\documentclass{aa}
\usepackage{txfonts}
\usepackage{graphicx}	
\usepackage{amsmath}	
\usepackage{amssymb}	
\usepackage{makecell}

\usepackage{natbib} 
\usepackage{color}
\bibpunct{(}{)}{;}{a}{}{,}
\usepackage{hyperref}
\definecolor{orange}{rgb}{0.99, 0.55, 0.01}

\begin{document}

   \title{The atmosphere of WASP-17b: Optical high-resolution transmission spectroscopy}


   \author{Sara Khalafinejad\inst{1,3,8} \and
          Michael Salz\inst{1} \and
          Patricio E. Cubillos\inst{2} \and George Zhou\inst{3} \and Carolina von Essen\inst{4} \and Tim-Oliver Husser\inst{5} \and
          Daniel D.R. Bayliss\inst{6} \and Mercedes L\'opez-Morales\inst{3} \and Stefan Dreizler\inst{5} \and J\"urgen H.M.M Schmitt\inst{1} \and Theresa L\"uftinger\inst{7}
          }
    
   \institute{Hamburg Observatory, Hamburg University, Gojenbergsweg 112, 21029 Hamburg, Germany,\\ \email{khalafinejad@mpia.de} \label{inst1}
        \and Space Research Institute, Austrian Academy of Sciences, Schmiedlstrasse 6, A-8042 Graz, Austria
        \and Harvard-Smithsonian Center for Astrophysics, 60 Garden Street, Cambridge, MA 01238, USA
        \and Stellar Astrophysics Centre, Aarhus University, Ny Munkegade 120, DK-8000 Aarhus C, Denmark
        \and Institute for Astrophysics, G\"ottingen University, Friedrich-Hund-Platz 1, 37077 G\"ottingen, Germany
        \and Department of Physics, University of Warwick, Coventry CV47AL, UK
        \and Institut f\"ur Astronomie, Universit\"at Wien, T\"urkenschanzstrasse 17, 1180 Wien, Austria
        \and Max Planck Institute for Astronomy, K\"onigstuhl 17, 69117 Heidelberg, Germany\\
             }


 
  \abstract{High-resolution transmission spectroscopy is a method for understanding
    the chemical and physical properties of upper exoplanetary
    atmospheres. Due to large absorption cross-sections, resonance lines
    of atomic sodium D-lines (at 5889.95~\AA\, and 5895.92~\AA\,) produce large transmission signals. 
    Our aim is to unveil the physical properties of \mbox{WASP-17b} through
    an accurate measurement of the sodium absorption
    in the transmission spectrum.
    We analyze 37 high-resolution spectra
    observed during a single transit of \mbox{WASP-17b} with the
    MIKE instrument on the 6.5 meter Magellan Telescopes.
    We exclude stellar flaring activity during the observations by analyzing   
    the temporal variations of H$_{\alpha}$ and Ca II infra-red triplet (IRT) lines. Then we obtain the excess absorption light curves in wavelength bands of 0.75, 1, 1.5 and 3 \AA\, around the center of each sodium line (i.e., the light curve approach). We model the effects
    of differential limb-darkening, and the changing planetary radial velocity
    on the light curves. We also analyze the sodium absorption directly in the
    transmission spectrum, which is obtained through dividing in-transit by
    out-of-transit spectra (i.e., the division approach). We then compare our measurements with a radiative transfer atmospheric model. 
    Our analysis results in a tentative detection of exoplanetary sodium: we measure the width and amplitude of the exoplanetary sodium feature to be $\sigma_{\mathrm{Na}}$ = (0.128 $\pm$ 0.078)~\AA\, and A$_{\mathrm{Na}}$ = (1.7 $\pm$ 0.9)\% in the excess light curve approach and $\sigma_{\mathrm{Na}}$ = (0.850 $\pm$ 0.034)~\AA\, and A$_{\mathrm{Na}}$ = (1.3 $\pm$ 0.6)\% in the division approach. By comparing our measurements with a simple atmospheric model, we retrieve an atmospheric temperature of 1550 $^{+170} _{-200}$ K and radius (at 0.1 bar) of 1.81 $\pm$ 0.02 R$_{\rm Jup}$ for WASP-17b.}
   
\maketitle
%
\section{Introduction}


Transmission spectroscopy, pioneered by \citet{Charbonneau2002}, is a very successful observational method to unveil the chemical and
physical properties of hot-Jupiter atmospheres. During an exoplanetary
transit, a fraction of the stellar light passes through the
planet's atmosphere, hence a fraction of this light is absorbed. As a result, the signatures of the chemical
composition of the atmosphere are imprinted on the transmitted
light at certain wavelengths, making the radius of the exoplanet to
appear larger at these wavelengths.



In the optical region the resonant doublets of sodium (Na), and
potassium (K), have large absorption cross
sections, hence, a larger absorption in the transmitted signal of these
lines is expected \citep{Seager2000, Fortney2010}. Thus, these lines have been
used as diagnosis of exoplanetary atmospheres
\citep[e.g.,][]{Vidal-Madjar2011, Huitson2012, Sing2012,
  Wyttenbach2017}. In addition, at optical wavelengths it is
possible to infer, for instance, water vapor \citep{Allart2017} and atmospheric hazes \citep[][]{Pont2008, Sing2016} and
molecular features of TiO \citep[e.g.,][]{Hoeijmakers2015, Sedaghati2017}.

Different types of instruments unveil different aspects of an exoplanetary atmosphere. For instance, using low-resolution
transmission spectroscopy, we can investigate hazes, clouds, Na, and 
K, in the deeper layers of exoplanetary atmospheres. On the other hand, by means of high-resolution transmission spectroscopy in narrow bands centered on Na doublets, we can access the information on upper atmospheric layers and the layers above the
hazes and clouds \citep{Kempton2014, Morley2015}.


Exoplanetary transit spectral observations are influenced by
different effects, mostly of stellar origins. These can be
flares, spots, differential limb-darkening and rotation
effects. A successful detection of the exoplanet's atmosphere requires a proper consideration of these effects.  We already investigated sodium in
\mbox{HD~189733b} as a prototypical target: \mbox{HD~189733A} is a very
active K-type star. After correcting for stellar flaring activity and stellar differential limb-darkening effects, we detected the signatures of sodium in the exoplanet \citep{Khalafinejad2017}. Here, we intend to apply a similar approach on transit observations of
\mbox{WASP-17b}, a hot Jupiter orbiting an inactive F-type star. 
Obtaining a comparative view of different targets helps to better understand the influence of stellar activity of different stellar types on transmission spectroscopy. 

In this study, we use the high-resolution spectral data of
\citet{Zhou2012}, who used a single transit observation of WASP-17b and claimed the detection of sodium by obtaining the excess light curve depth in passbands of 1.2~\AA\, to 1.8~\AA\, (with steps of 0.1\AA) around the center of sodium lines and measured the depths by fitting a simple light curve model.
We analyze the sodium absorption in
\mbox{WASP-17b} with a new approach, we intend to improve and complete the previous analysis by \citet{Zhou2012}:
We rely on their extraction of the raw excess light curve and go beyond the previous work, first by accounting
for the stellar limb-darkening effect and changes in the radial velocity of the planet in the excess light curves. The second consideration is that, in addition to the detection of sodium, we constrain the physical characteristics of WASP-17b by modeling the sodium transmission spectrum of the planet. 
Finally, we additionally investigate
the behavior of the H$_{\alpha}$ and \ion{Ca}{ii} infra-red triplet (IRT) lines during the transit to inspect
the stellar activity and possibly the exoplanet's atmospheric hydrogen absorption from the upper atmosphere of this highly inflated \citep[][]{Anderson2010} exoplanet.

The structure of this paper is as follows: 
In section \ref{sec:WASP-17}, we introduce the hot-Jupiter WASP-17b. Then in Section \ref{sec:obs} we describe the observation, and the steps of data reduction.
In Section \ref{sec:analysis_activity_wasp17}, we investigate the stellar activity through H$_{\alpha}$ and \ion{Ca}{ii} IRT. Section \ref{sec:analysis_excess_wasp17} focuses on detection of exoplanetary atmospheric sodium using the excess light curves, and Section \ref{sec:Division} presents the method and results for extraction of the transmission spectrum in the region of sodium lines. In Section \ref{sec:discussion} we discuss the results and compare the observations with an atmospheric model. Finally in Section \ref{sec:conc} we summarize our conclusions.

\section{The system WASP-17}
\label{sec:WASP-17}
WASP-17b is an inflated hot-Jupiter, with a mass of 0.49 M$_{\rm Jup}$, a radius of 1.99 R$_{\rm Jup}$ and an orbital period of $\sim$3.7 days \citep[]{Anderson2010} and it is in a retrograde motion
\citep[]{Triaud2010,Bayliss2010}. Its host star is of spectral type
F6V, with a magnitude of V = 11.6 \citep[][]{Hog2000}. The
parameters of the system used in this work are
summarized in Table \ref{tbl:orbital parameters}. \mbox{WASP-17b} has
a very low density, $\sim 0.06 \rho_{Jup}$, and a high equilibrium
temperature, \mbox{$\sim$1800 K} \citep[]{Anderson2011}. Thus, it is
expected to have a very large atmospheric scale height. In consequence, this hot Jupiter has become one of the few very well studied targets for
atmospheric characterization with transmission spectroscopy.
For instance, \citet{Wood2011} used medium-resolution \mbox{($\mathcal{R}$ $\sim$
  12\ 500)} observations with the GIRAFFE fiber-fed spectrograph at the
VLT and reported a sodium detection with an excess absorption of (1.46
$\pm$ 0.17)\%, (0.55 $\pm$ 0.13)\% and (0.49$\pm$ 0.09)\% in passbands of
0.75 \AA\,, 1.5 \AA\, and 3 \AA\,, respectively. \citet{Zhou2012} used
the Magellan Inamori Kyocera Echelle (MIKE) spectrograph \mbox{($\mathcal{R}$
  $\sim$ 48\ 000)} on the Magellan Telescopes for the same purpose. By
applying the transmission spectroscopy technique in narrow bands they
detected a (0.58 $\pm$ 0.13)\% signal at the core of sodium D-lines (D$_{2}$ at 5889.95~\AA\, and D$_{1}$ at 5895.92~\AA)
in a passband of 1.5 \AA. 

Later, low-resolution broad-band observations
in the optical and IR region were performed. \citet{Bento2014} used
multi-color broad-band photometry with SDSS 'u','g','r' (covering
wavelengths from 325 to 690 nm) and compared to the g-band they found evidence for increased absorption in the r-filter where the sodium feature is located. Using the Hubble Space Telescope (HST) WFC3 instrument, \citet{Mandell2013} analyzed low-resolution transmission
spectroscopy in the IR region (1.1 -1.7 $\mu$m). Their analysis of the
band-integrated time series suggests water absorption and the presence
of haze in the atmosphere of \mbox{WASP-17b}. \citet{NortmannThesis}
used FORS2 at the VLT, and after dealing with instrumental
systematics, derived the optical transmission spectrum of
\mbox{WASP-17b} in the region between 800 and 1000 nm. Their transmission spectrum hints at strong absorber in the bluer wavelength region.
In addition, based on their tested theoretical models, tentative signatures of a TiO and VO or a potassium and water atmosphere have been observed, but these models cannot fully explain their observations.
\citet{Sing2016}
used HST (STIS and WFC3) and Spitzer observations, and obtained optical to mid-IR transmission spectrum ($\sim 0.3\mu$m - 5$\mu$m) of
a sample of hot-Jupiters including WASP-17b. In their measurements, the atmospheric absorption in WASP-17b, at $\sim$5900~\AA\, (wavelength of sodium) with the bin size of 5~\AA\,, is about $(0.33 \pm 0.18)$\%.
\citet{Sedaghati2016} analyzed
low-resolution ($\mathcal{R}$ $\sim 2000$) transit observations of FORS2 and obtained the
broad-band transmission spectrum of WASP-17b with a bin size of 100~\AA\, in the wavelength range of 5700 to 8000 \AA, where they detected a cloud-free atmosphere and
potassium absorption with a 3$\sigma$ confidence
level. In their study the sodium feature was not detected neither in 100~\AA\, nor in 50~\AA\, bin size. Finally, \citet{Heng2016}'s work also suggested a nearly cloud-free atmosphere at visible wavelengths.

\begin{table*}
\centering
\label{tbl:orbital parameters}
\begin{tabular}{ l l l }
\hline\hline
Parameter & Symbol & Value  \\
\hline
        Mid-transit time               & $T_{0}$ (HJD$_{\text{UTC}}$) & 2454577.85806 $\pm$ 0.00027    \\
        Orbital Period                 & $P$ (days)             & 3.7354380 $\pm$ 0.0000068      \\
        Transit duration               & $T_{14}$ (days)  &    0.1830 $\pm$ 0.0017 \\
        Ingress/egress duration        & $T_{12}$ =T$_{34}$ (days)   &  0.0247 $\pm$ 0.0017  \\
        Orbital inclination            & $i$ ($^{\circ}$)      & 86.83 $\pm$  0.62      \\
        Semi-major axis                & $a$ (au)          & 0.05150 $\pm$ 0.00034        \\
        Planet mass                    & $M_{\rm p}$ (M$_{\rm Jup}$)  & 0.486 $\pm$ 0.032  \\
        Planet radius                  & $R_{\rm p}$ (R$_{\rm Jup}$)  & 1.991 $\pm$ 0.081   \\
        Planet surface gravity         & log $g_{\rm P}$ (cgs)         & 2.448 $\pm$ 0.042 \\
        Stellar effective temperature  & $T_{\rm eff}$ (K)    & 6650 $\pm$ 80           \\
        Stellar surface gravity        & log $g$ (dex)          & 4.161 $\pm$ 0.026       \\
        Metallicity                    & $[\text{Fe/H}]$      &  -0.19 $\pm$ 0.09       \\
        Stellar radius                 & $R_{\rm S}$ ($R_{\odot}$)    & 1.572 $\pm$ 0.056    \\ 
        Projected stellar rotation velocity       & \textit{v}sin(\textit{i}) (kms$^{-1}$)         & 10.05 $\pm$ 0.88       \\
        Planet to star area ratio    & ($R_{\rm P}$/$R_{\rm S}$)$^{2}$          & 0.01696  $\pm$ 0.00026      \\ 
\hline
\end{tabular}
\caption {Adopted values for the orbital and physical parameters of WASP-17 used in the ecxess light curve approach in this work (\ref{sec:analysis_excess_wasp17}). For consistency with \citet{Zhou2012}, all values are taken from \citet{Anderson2011}}.
\end{table*}

\section{Observations and data reduction}
\label{sec:obs}
On the night of 2010 May 11, one single transit of the hot Jupiter,
\mbox{WASP-17b}, was observed with the MIKE spectrograph, mounted on the 6.5m Magellan II (Clay)
Telescope. During the observation 37 spectra, with an
exposure time of 600 seconds each were obtained. In addition, a slit widths of
0.35 arcsec was chosen, resulting in a spectral resolution of
\mbox{$\mathcal{R} \sim 48,000$} in the wavelength
range of 5000 to 9500 \AA. The spectra have a pixel scale of $\sim$
0.04 \AA\, and a resolution element of about 0.12 \AA. In the sodium
doublet region, the S/N of our spectra is about 80 per pixel. Two of the
spectra with poor signal-to-noise (S/N < 60) and high airmass were
discarded. These correspond to the first and last exposures. The details 
of the observations, along with the initial data reduction of the echelle 
spectra, are fully discussed by \citet{Bayliss2010}.

\subsection{Spectral data reduction in Na, H$_{\alpha}$ and \ion{Ca}{ii} IRT}
\label{sec:data_reduction}

The spectra were analysed in wavelength bands with a width of a few hundred Angstroms in three regions: Na (5800-6100~\AA\,), H$_{\alpha}$ (6500-6700~\AA\,), and \ion{Ca}{ii} IRT (8400-8700~\AA\,). Data preparation before performing the main analysis mainly consists of spectral normalization, spectral alignment and telluric line removal.
The data reduction in the Na doublet, H$_{\alpha}$ and \ion{Ca}{ii} IRT regions follows a similar procedure. For the spectral normalization and removal of telluric features in the Na region we rely on the data reduction by \citet{Zhou2012}. Here, we accurately align the spectra in the Na wavelength region, in addition we perform a complete reduction around the H$_{\alpha}$ and \ion{Ca}{ii} IRT lines.

\begin{figure}
    \includegraphics[width = \columnwidth]{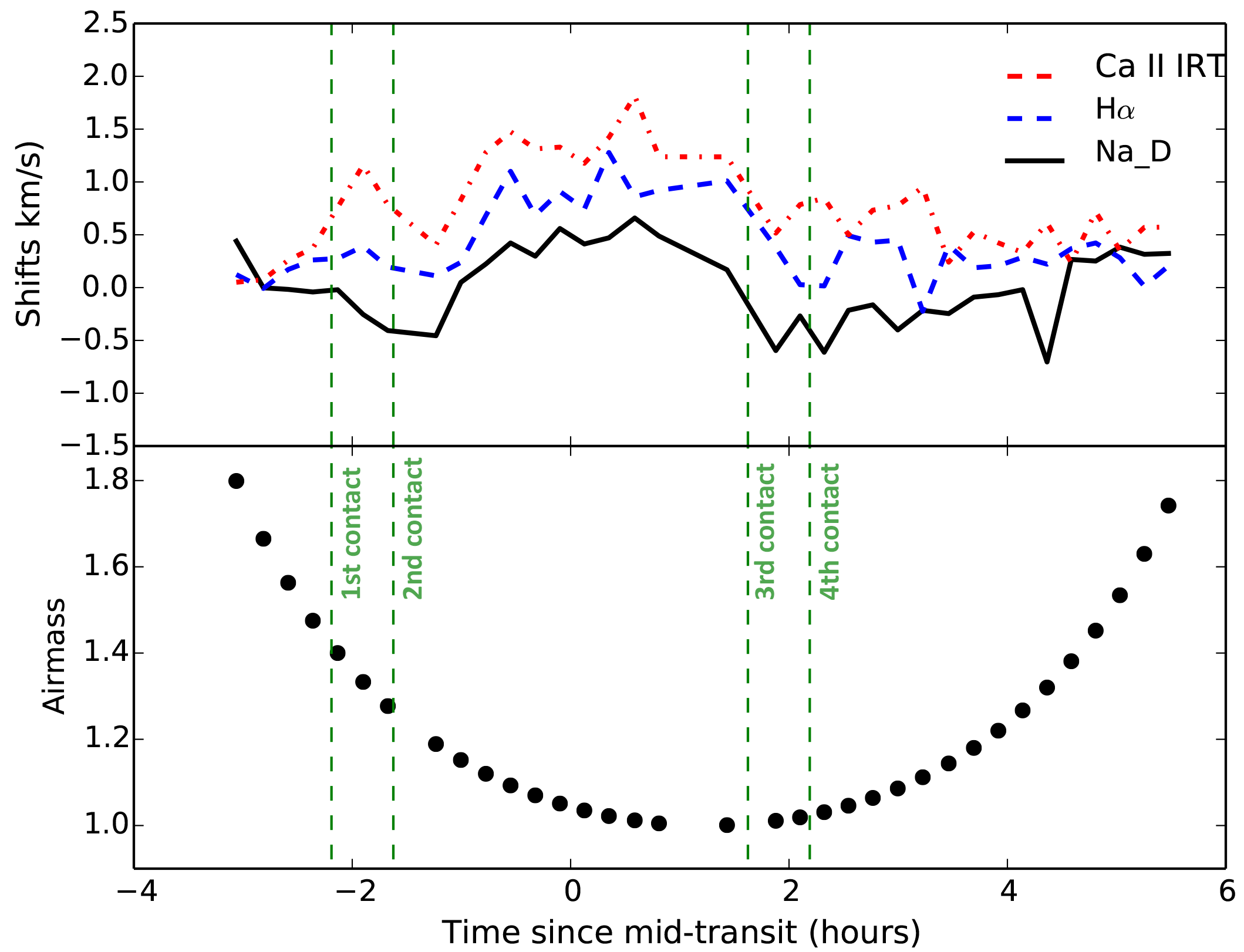}
    \caption{\textbf{Top:} Spectral misalignments in three regions around the Na doublet, H$\alpha$ and Ca II IRT line. \textbf{Bottom:} Airmass value for each exposure (two of the exposures with airmass larger than 1.8 are not shown).}
    \label{fig:AM_shift}
\end{figure}

We shift the spectra into the stellar rest frame. To correct the misalignment between exposures, we select 8,7, and 14 stellar spectral lines in the vicinity of the Na doublets, H$_{\alpha}$, and \ion{Ca}{ii} IRT lines respectively and fit Gaussians to determine the line centers. Each spectrum is shifted by the mean offset of the line centers with reference to the first observation. The resulting misalignments in each region are shown as radial-velocity (RV) shifts in Fig.~\ref{fig:AM_shift} (top). All regions show a similar pattern, with a maximum difference of $\sim$1~km\,s$^{-1}$ between exposures.
We then correct cosmics via linear interpolation over the affected spectral ranges and interpolate all spectra onto a common wavelength grid, increasing the sampling by a factor of four to minimize interpolation errors. 5th-order polynomials are used to normalize the continuum.

The airmass value of each exposure is also shown in Figure \ref{fig:AM_shift} (bottom). Changes of airmass cause variations in the telluric features (mainly water at these wavelengths) with a similar trend.
To remove the tellurics in the sodium region, the method by \citet{Zhou2012} is followed: The observations of the rapidly rotating B star, HD~129116, are used as a template. For each exposure, this telluric template is scaled to fit the prominent telluric water lines and then the spectra are divided by the template.
As explained in \citet{Zhou2012}, no telluric sodium is observed.
The spectra of WASP-17 are affected by interstellar sodium lines which has wavelength shifts of about 1~\AA\, with respect to the stellar lines.

\begin{figure*}
    \centering
    \includegraphics[width=\textwidth]{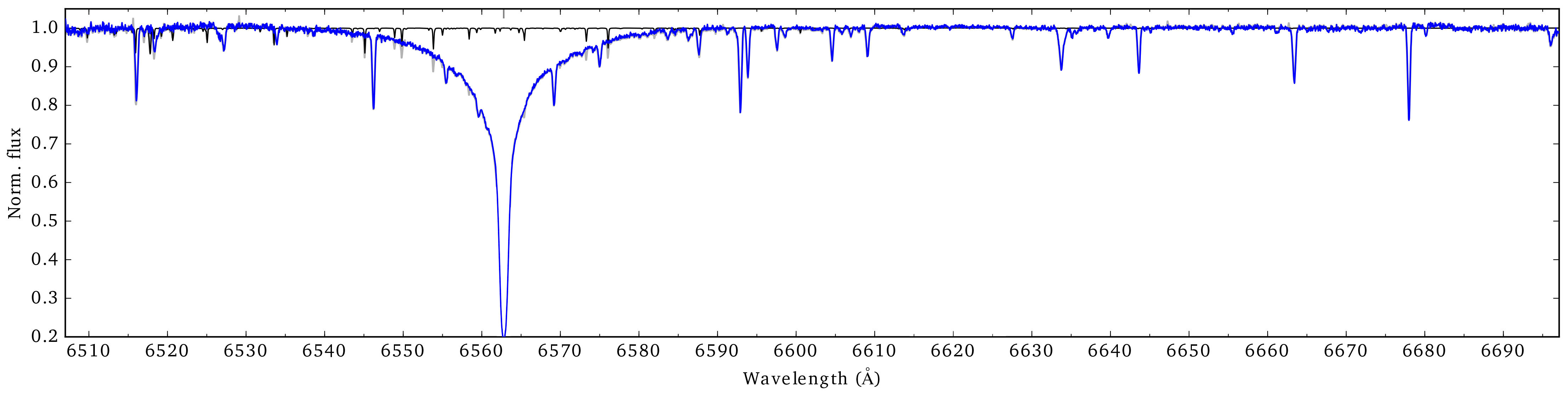}\\
    \includegraphics[width=\textwidth]{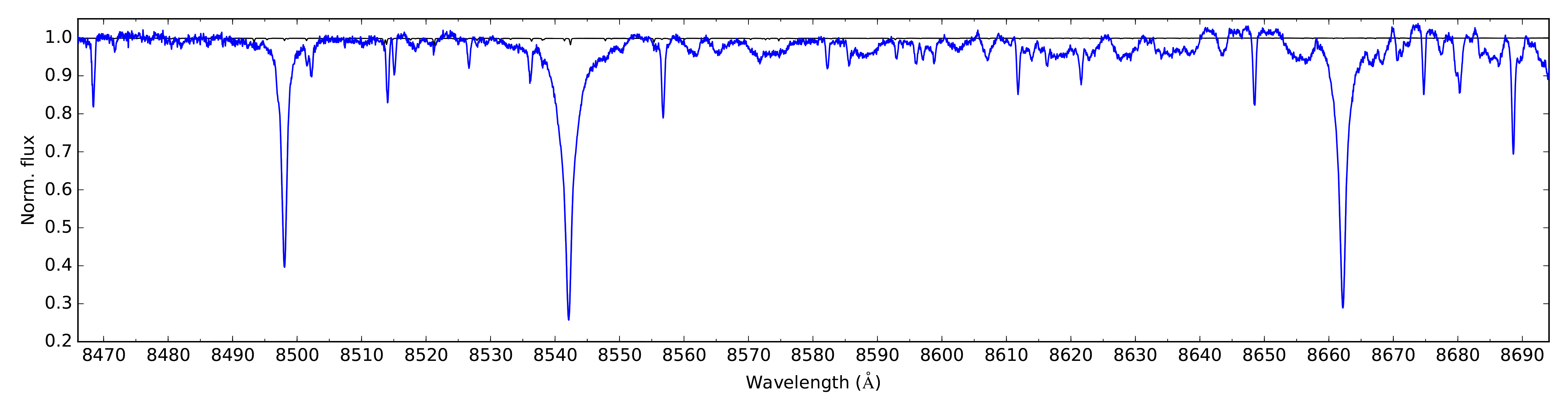}
    \caption{Average H$\alpha$ and \ion{Ca}{ii} IRT spectra (blue). 
             Black lines show a telluric transmission spectrum used
             for the telluric correction in the H$\alpha$ region.
             The uncorrected spectrum is shown in gray in the H$\alpha$
             panel.
             }
    \label{fig:Halpha-CaIRT-region}
\end{figure*}

The removal of the Earth's atmospheric features in the H$\alpha$ and \ion{Ca}{ii} IRT regions is a bit different. Telluric lines are stronger in the H$\alpha$ range and we devise an approximate approach to remove them. A high resolution telluric transmission spectrum of \citet{Moehler2014} for an airmass of 1.5 is fitted to the average stellar spectrum. This fitting procedure has three free parameters: The stellar radial velocity, the amplitude of the telluric lines, and the width of a Gaussian with which the telluric spectrum is convolved to model the instrumental resolution. In a second step, the observation run is split into five sections and for each of the five mean spectra, we again fit the telluric spectrum by only adjusting the amplitude of the telluric lines. During the observation night, this amplitude follows a trend that resembles the airmass trend, and we fit a 2nd order polynomial to the amplitude evolution. Each spectrum is then divided by the shifted and convolved telluric reference spectrum with the amplitude derived from the polynomial. Our visual inspection shows that, in the individual spectra, strong telluric lines are effectively reduced at the 90\% level. Figure~\ref{fig:Halpha-CaIRT-region} shows the average spectrum of WASP-17 in the H$_{\alpha}$ and \ion{Ca}{ii} IRT regions with the telluric transmission spectrum and the uncorrected spectrum for the H$\alpha$ range. The telluric correction has little impact on the resulting H$\alpha$ equivalent widths (see Section \ref{sec:activity}).

\section{Data analysis: stellar activity}
\label{sec:analysis_activity_wasp17}
Using high-resolution transit spectra, we have two possible approaches to detect the exoplanetary sodium embedded inside the stellar spectrum: One is to integrate in narrow passbands inside the stellar absorption lines (with a reference band in the continuum) and investigate the additional contribution of the planetary atmosphere to the obscuring of the stellar light during the transit (excess light curve approach). The other is to directly obtain the exoplanetary transmission spectrum by dividing the in-transit spectra by the out-of-transit spectra (division approach). But first we identify possible signatures of stellar activity on the spectra.
In this section we investigate the stellar activity and then, in the next sections, we perform the transmission spectroscopy both through excess light curve and division approaches.

\subsection{Investigations of stellar activity}
\label{sec:activity}

Stellar spots \citep[e.g.,][]{Oshagh2013,Czesla2009}, plage regions \citep[e.g.,][]{Oshagh2014} and flaring activity \citep[e.g.,][]{Klocova2017}, are among the main sources of false signals in the interpretation of transmission spectra.

Small bumps or dips in the photometric transit light curves and deformations of the high-resolution spectral line shapes can be signatures of stellar variability. Investigations of spots through study of spectral line deformations are easier for very fast rotating stars \citep[e.g.,][]{Wolter2005,Reiners2012}.
In this work, we do not have simultaneous photometric observations and the star is a slow rotator, hence study of stellar spots is not possible. However we are still able to perform an in-depth study of flaring events.

Chromospheric lines such as H$_{\alpha}$ (at 6563~\AA) and \ion{Ca}{ii} IRT (at 8498, 8542, 8662~\AA) are indicators of stellar activity \citep[e.g.,][]{Cincunegui2007, Martinez2010, Chmielewski2000, Andretta2005, Busa2007, Klocova2017}, but also other stellar lines such as the sodium D-lines can be affected by stellar activity \citep[e.g.,][]{Cessateur2010}. 
Here, we estimate by how much a planetary signal in the sodium D-lines can be affected by stellar activity.

Additionally, the H$_{\alpha}$ line may indicate signatures of hydrogen escape from the upper atmosphere of hot gas planets \citep{Cauley2016, Cauley2017}. Due to its low mean density and a high irradiation level, WASP-17\,b is expected to host a strongly evaporating atmosphere \citep{Bourrier2015, Salz2016}. Therefore, it is reasonable to search for H$\alpha$ absorption features around the planetary transit. 

We compute the equivalent width (EW) of the individual ``transmission'' spectra in the H$_{\alpha}$ and all three \ion{Ca}{ii} IRT lines following \citet{Cauley2017}. Each spectrum is divided by the total mean
spectrum and then integrated over the central $\pm$50 km s$^{-1}$. The EW of the three IRT lines is then averaged. Errors are derived from the variation in adjacent $\pm$500 to 1000 km s$^{-1}$ bands. This slightly underestimates the error in the line cores and the impact of red-noise, therefore, we follow \citet{Czesla2017} to derive a mean
error for the EW curves and scale our values accordingly.

\begin{figure*}
    \centering
    \includegraphics[width=0.8\textwidth]{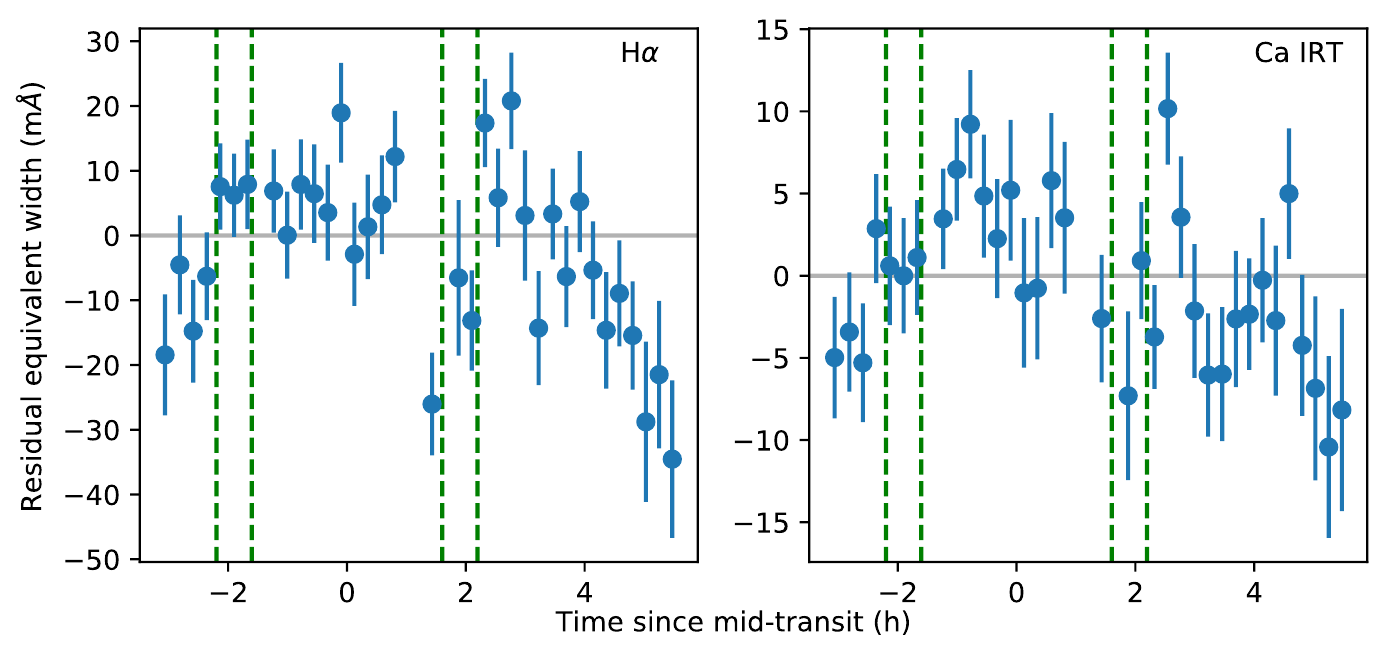}
    \caption{Time evolution of the equivalent widths of the H$_{\alpha}$ line and the mean of the three Ca II IRT lines. The contact points of the transit are indicated by vertical dashed green lines. The horizontal line shows the reference value of zero. Both lines show some activity evolution over the night, but no stellar flares.}
    \label{fig:EW}
\end{figure*}

The time evolution of the equivalent widths is shown in Fig.~\ref{fig:EW}. Both EWs evolve similarly: except for the planetary egress, WASP-17 shows a slightly increased activity level during the planetary transit compared to
the out-of-transit level. This is also seen in the average in-transit transmission spectrum of the H$_{\alpha}$ and IRT regions (see Fig.~\ref{fig:TS_Halpha_CaIIIRT}).
At egress both EW light curves show a dip, which can be caused by stellar variability or by some absorption feature of planetary origin. However, the signal-to-noise is not sufficient to investigate this
any further. The H$_{\alpha}$ EW of WASP-17 changes by about 20~m\AA{} over the transit duration, which is slightly stronger than that of HD 189733 \citep[see][]{Cauley2017}. The EW of a planetary H$_{\alpha}$ absorption feature should be smaller than the observed activity variation, which is equivalent to an absorption level 0.9\% over the integration band. For the IRT we exlude 8~m\AA{} corresponding to 0.3\% absorption. 
WASP-17 is among several other systems with non-detections of H$_{\alpha}$ in-transit absorption (HD 149026, HD 147506, KELT-3 b, and GJ 436 b \citealt{Jensen2012, Cauley2017b}). Currently, HD~189733 \citep{Jensen2012, Cauley2017} and KELT-9b \citep{Yan2018} remain the systems with
possible H$_{\alpha}$ absorption caused by the planet’s evaporating atmosphere.

In the study of \citet{Klocova2017} a moderate flare on the K-type host star HD~189733 affected the time evolution in all studied spectral lines with a similar time behaviour.
\citet{Khalafinejad2017} showed that the sodium lines were affected by the same flare, but on a level about a factor of 10 weaker than the H$_{\alpha}$ line. 
In WASP-17, the activity seen in the H$_{\alpha}$ region of 20 m\AA\ could affect the Na region on a 2~m\AA\ level, if we assume a similar scaling as in HD~189733. This is about half of the level of the observed absorption signal in the Na region (see below). However, the activity is higher during the transit and should not cause an artificial absorption signal, but rather decrease the planetary sodium signal. We do not observe a clear stellar flare and, thus, do not attempt to correct the impact of stellar activity variations in the sodium region, but we do note that the observed absorption level may have been lowered by an increased in-transit activity level of the host star.

\section{Analysis and results: Excess light curve approach}
\label{sec:analysis_excess_wasp17}
\subsection{Extraction and modeling of the excess light curve}

The extraction of the raw excess light curves in the sodium doublet region is performed by \citet{Zhou2012}, who use integration passbands of 0.75, 1, 1.5 and 3 \AA\, centered on the core of each sodium line 
and select the interstellar sodium next to D1 (Na line with the longer wavelength) as a reference in the flux integration. For consistency with \citet[][]{Zhou2012}, we use the same integration and reference bands.
During the transit the radial velocity of the planet changes between $-18$ to $+18$ kms$^{-1}$, which results in a Doppler shift of up to $\pm$ 0.35~\AA. Thus, with a passband of 0.75 \AA\, we can still be sure that the exoplanetary feature is still located inside the integration band.
Here we use a different approach in measuring the exoplanetary signal in the excess light curves.
We make a combined model of differential LD and changing RV  models to fit the excess light curves; the detailed explanations of these two model components are presented by \citealt{Khalafinejad2017}.
To increase the signal-to-noise we use average of both sodium D-lines and do not treat the lines individually.

There are some effects that inevitably influence transit spectra or excess light curves. To accurately model the excess light curves, we take the following effects into account and consider each as a model component of the final model, which is introduced in Section \ref{sec:Final_model}.

\subsubsection{Differential Limb-darkening (component A)}

\begin{table*}
\small{
\begin{center}
\begin{tabular}{ c | l c c c c c}
\hline\hline
\textbf{$\lambda$-range (\AA\,)} & \makecell{\textbf{D2}: \\ \textbf{D1}:} & \makecell{5888.52-5889.27 \\ 5894.48-5895.23} & \makecell{5888.40-5889.40 \\ 5894.36-5895.36} & \makecell{5888.15-5889.65 \\ 5894.11-5895.61} & \makecell{5887.40-5890.40 \\ 5893.36-5896.36} & 5895.50-5898.00 \\
\hline
\textbf{Average}        &  & 0.75 \AA\, core & 1 \AA\, core        & 1.5 \AA\, core     & 3 \AA\, core         & ref. \\
\textbf{u$_{1}$,u$_{2}$}   &      & 0.3651 , 0.1447   & 0.3850, 0.1589    & 0.3965 , 0.1802     & 0.3979 , 0.2144 & 0.4009 , 0.2306 \\
\hline

\end{tabular}
\end{center}
}
\caption {\label{tbl:LDs} Limb-darkening coefficients (u$_{1}$,u$_{2}$) for the average of
  sodium D$_{2}$ and D$_{1}$ integration bands (Core) and
  reference (Ref.) bands with the specified wavelength ranges in the table below. Errors of the limb-darkening coefficients are in all cases on the
  order of 10$^{-3}$-10$^{-4}$.}
\end{table*}

Stellar limb-darkening depends on wavelength \citep[e.g.,][]{Czesla2015}. Thus dividing the flux in the integration band by the reference band, is similar to dividing two light curves with two different limb-darkening coefficients by each other. Thus, this division affects the shape of the excess light curve and needs to be corrected for, specifically in late type stars where the effect is more prominent.
We calculate the limb-darkening coefficients in each passband using the intensity profiles of the PHOENIX model \citep{Hauschildt1999, Husser2013} and the quadratic limb-darkening law \citep[][]{Kopal1950}. More details are explained by \citet{Khalafinejad2017}; the results are shown in Table \ref{tbl:LDs}. Since the excess light curves are already averaged out between D1 and D2, we use the average of the coefficients. The signal-to-noise of our data is not enough to consider LD coefficients as fitting parameters, hence this model is considered to be constant and to have no fitting parameters. We note that, changing this constant model component by 50\%, alters the final measured sodium absorption within the 1$\sigma$ error-bar.

\subsubsection{Planetary radial velocity (component B)}

The exoplanetary spectral lines move inside the stellar lines due to the Doppler shift caused by the exoplanetary orbital motion. This results in an apparent reduction of the depth of excess light curve at the mid-transit time \citep{Khalafinejad2017, Albrecht2008Thes}.

For considering the exoplanetary atmospheric absorption effect in this work, the exoplanetary sodium line is considered as a Gaussian convolved inside the stellar sodium line. This Gaussian moves inside the stellar sodium line with an offset proportional to the change of the radial velocity of the exoplanet by the orbital motion \citep[see equation 8 in][]{Khalafinejad2017}. At the same time, the moving Gaussian causes the dip in the light curve and a bump near the mid-transit time. The fitting parameters of this components are the width (Gaussian $\sigma_{Na}$) and the depth (Gaussian A$_{Na}$) of the exoplanetary Gaussian profile. Since our raw excess light-curve is the average of both sodium D1 and D2, we model the effect on both stellar sodium lines and then obtain the averages of sigma and amplitude of the Gaussian feature.

\subsubsection{Neglected Effects}
According to \citet{Triaud2010} and \citet{Bayliss2010}, WASP-17 has a large spin-orbit misalignment angle ($\sim 150^{\circ}$), and the Rossiter-McLaughlin (RM) effect does not produce a symmetric RV curve. Therefore, the variations in line shape during the transit do not completely cancel out. Based on Figure 6 in \citet[][]{Anderson2010}, the difference between the amplitude of the curve in the blue- and red-shifted parts of the star is about 100 ms$^{-1}$, which results in a 0.002~\AA\, change of a spectral line position. This value is at least one order of magnitude smaller than what we expect for the width of the exoplanetary sodium feature. Thus we ignore the RM effect in this analysis. We note that we evaluate the influence of the RM effect on measurements of the line centers in the alignment of spectra in Section \ref{sec:extract_TS}. However, the maximum radial velocity caused by the RM effect is smaller than the precision of the alignment, thus no evidence of influence of this effect could be detected.

In addition, as mentioned in Section \ref{sec:activity}, we cannot take into account the effects of possible spots and plages in this data set. In any case, as also confirmed through our investigation of the activity indicating lines, we do not expect pronounced activity features on this F-type star.

\subsubsection{Final model}
\label{sec:Final_model}
A combination of the LD and the RV models constitutes the main model, we also consider a normalization constant (offset) as the third model component (C) and consider the model, M, as a function of time, t:\\ 

\begin{equation}\
\begin{array}{ll}  
\label{eq:model}
M(t) & = LD(t) \times RV(t) + offset\\
& \equiv A \times B + C\\
\end{array}
\end{equation}

The fitting parameters in the model, in Eq. \ref{eq:model}, are the $\sigma_{Na}$, A$_{Na}$ and the offset.
For exploring the best-fit parameters
and their associated uncertainties we apply a Markov Chain Monte
Carlo (MCMC) analysis, using the affine invariant ensemble sampler
\emph{emcee} \citep{Foreman2013}. We employ 20 walkers, with 100
chains each, where the initial positions are synthesized from a
Gaussian distribution around our best estimates. All the free parameters have uniform priors imposed. For $\sigma_{Na}$ and A$_{Na}$ we set a lower bound of 0 and upper bound of 0.3.
We allow a burn-in phase of $\sim 50\%$ of the total chain length, beyond which the MCMC
is converged. The posterior probability distribution is then
calculated from the latter 50\% of the chain (see Section \ref{sec:best-fit parameters} for the results).

\subsection{Best-fit parameters and uncertainties of the excess light curve models}
\label{sec:best-fit parameters}

\begin{table}
\centering
\begin{tabular}{c c  c  c }
\hline\hline
Passband &  $\sigma_{\mathrm{Na}}$ (\AA\,) &  A$_{\mathrm{Na}}$ & offset\\
\hline

0.75\AA\, & 0.128 $\pm$ 0.078  & 0.017 $\pm$ 0.009   & 0.007 $\pm$ 0.001   \\
1\AA\, & 0.054 $\pm$ 0.040  & 0.034 $\pm$ 0.025   & 0.005 $\pm$ 0.001  \\
1.5\AA\, & 0.077 $\pm$ 0.074  & 0.031 $\pm$ 0.029  & 0.002 $\pm$ 0.001  \\
3\AA\, & 0.036 $\pm$ 0.061  & 0.054 $\pm$ 0.067  & 0.001 $\pm$ 0.001   \\
\hline

\end{tabular}
\caption {\label{tbl:best-fit-values} Best-fit values for the model parameters obtained from the
  Na D$_{1}$ and Na D$_{2}$ excess light curves in
  the three integration bands. As a reminder, here \boldmath$\sigma_{\mathrm{Na}}$ is width of the exoplanetary Gaussian profile, \textbf{A$_{\mathrm{Na}}$} is the amplitude of the exoplanetary Gaussian profile, and \textbf{offset} is the normalization constant.}
\end{table}

\begin{figure*}
    \centering
    \includegraphics[width = 0.75\textwidth]{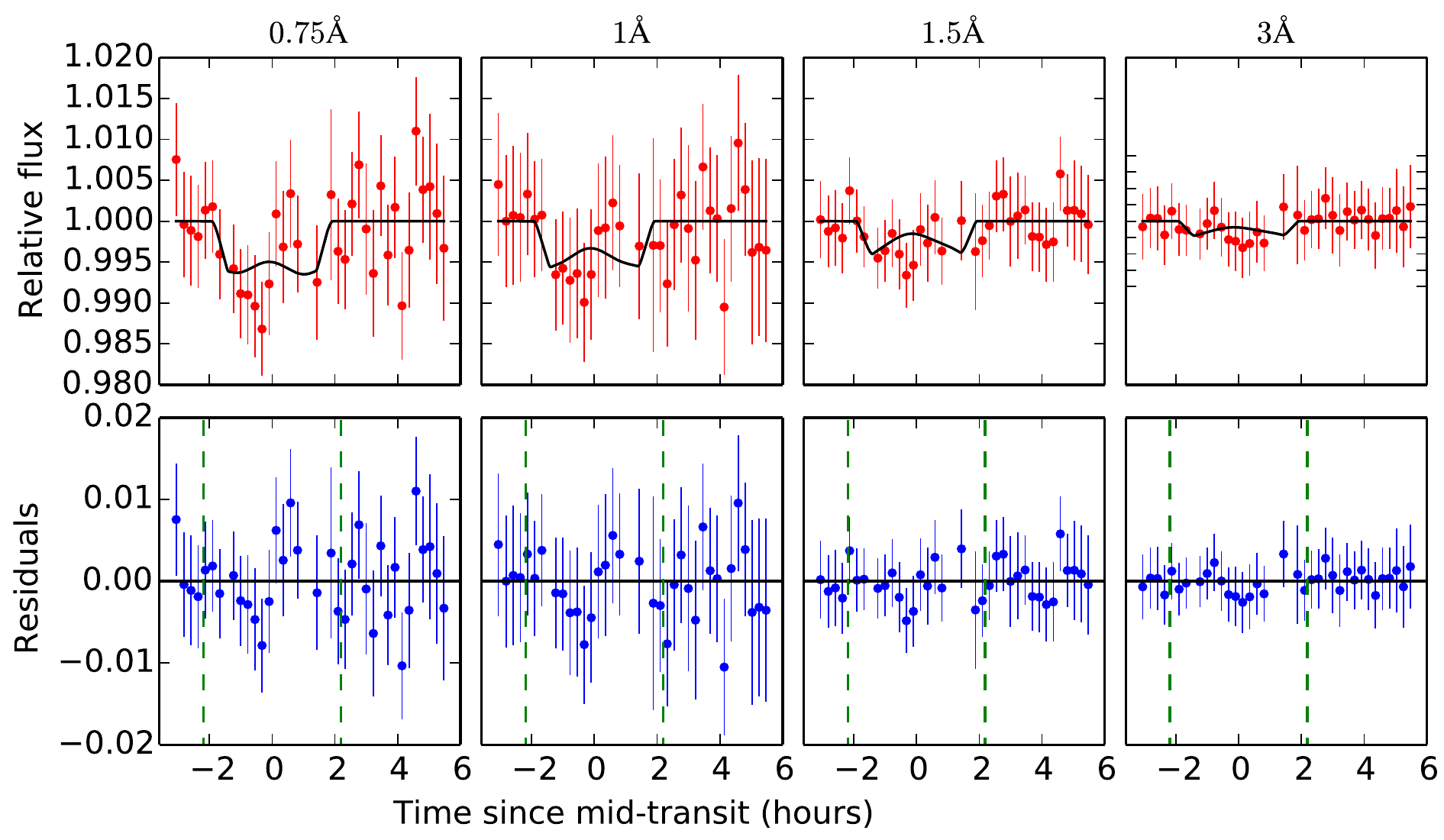}
    \caption{\textbf{Top panels:} Best-fit models plotted over the raw excess light-curves for four passbands. \textbf{Bottom panels:} The residuals for each passband.}
    \label{fig:best-res}
\end{figure*}

The best-fit model of each excess light curves are shown in the top panels of Figure \ref{fig:best-res} and the values of the parameters as well as their uncertainties are provided in Table \ref{tbl:best-fit-values}. The posterior distributions of the fitting parameters and the correlation plots are shown in the appendix Figure \ref{fig:corner} for the 1.5~\AA\, excess light curve. The bottom panels in Figure \ref{fig:best-res} show residuals after subtracting the data from the model. The excess light curves have a large noise level and thus the atmospheric absorption measurements have large uncertainties (see Table \ref{tbl:best-fit-values}). As Table \ref{tbl:best-fit-values} shows, the significance of the signal is about $2\sigma$ in our narrowest integration band and the significance level reduces by the increase of the passband. At 1.5~\AA\, there is hardly a signal and the 3~\AA\, passband results in a null outcome. At narrower passbands the contrast between the planetary and stellar flux contributions is smaller, thus a larger signal is expected.
The measured exoplanetary signal with this approach is compared with planetary atmospheric models in Section \ref{sec:atm_model_LC}.  

\section{Analysis and results: Division approach}
\label{sec:Division}
\subsection{Extraction of the exoplanetry spectrum through the division approach}
\label{sec:extract_TS}

\begin{figure*}
    \centering
    \includegraphics[width = 0.75\textwidth]{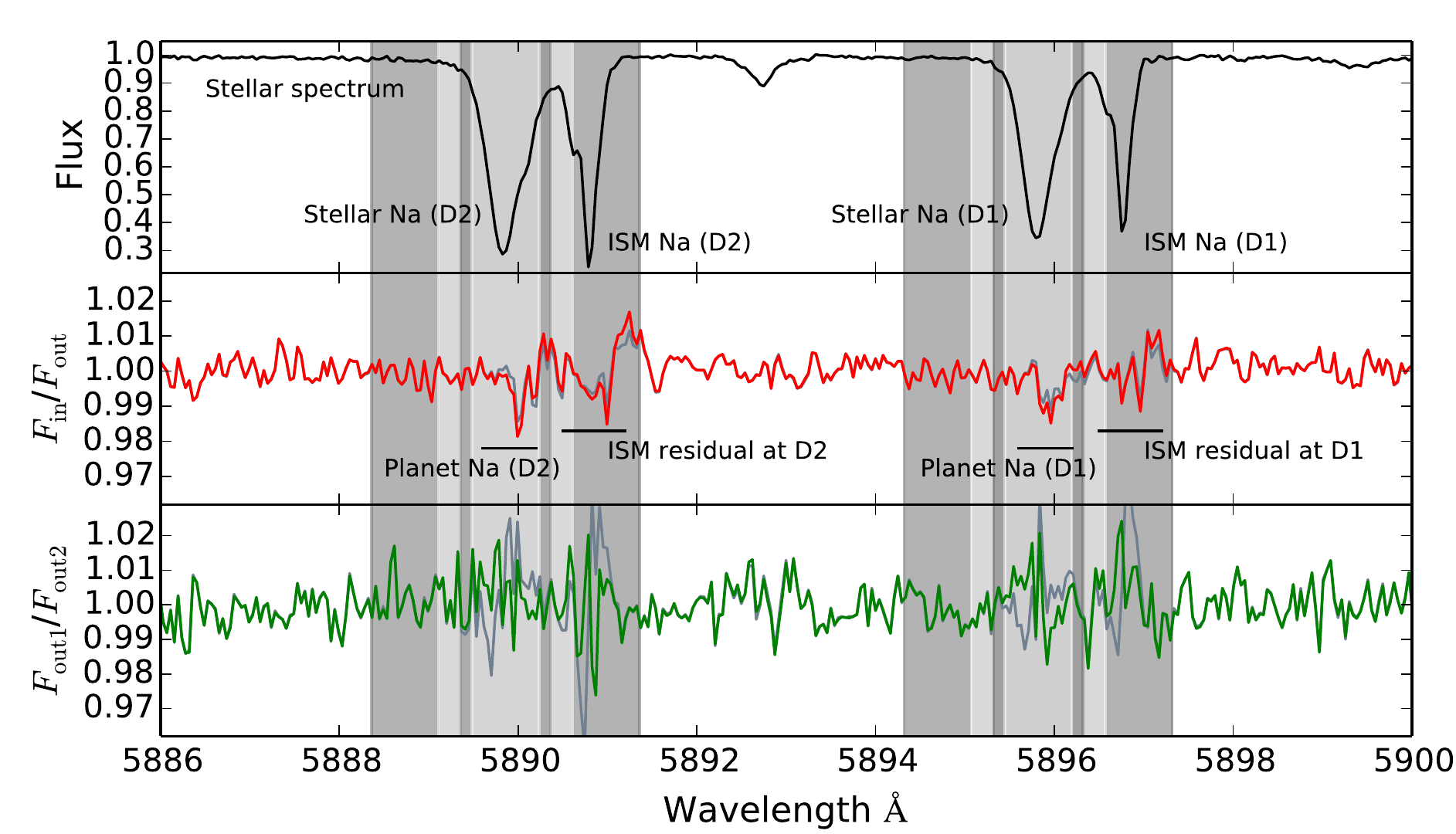}
    \caption{Transmission spectrum obtained by division approach. The black profile in the \textbf{first panel} is the stellar spectrum. The red profile in the \textbf{second panel} is the division of the in-transit by the out-of-transit spectra, corrected for the exoplanetary RV shifts. The dark gray profile beneath the red profile is the division after applying the trial manual shifts. The dark gray profile in the \text{third panel} is the division of two sets of out-of-transit spectra. The green profile is the same division shifted with respect to the other by 0.004~\AA. Finally, the shaded regions, represent the integration bands at 0.75, 1, 1.5, and 3 \AA.}
    \label{fig:inout}
\end{figure*}

In transmission spectroscopy we compare the spectra taken when the exoplanet is outside the transit with those taken during the transit. Only the in-transit data contain the planet's atmospheric signatures, hence subtracting the out-of-transit spectrum from them, can reveal the exoplanetary features within the residuals.
The exoplanetary transmission spectrum is given by ($F_{\rm in} - F_{\rm out}) / F_{\rm out}$. 
In this simple equation, the average of all in-transit spectra is considered as $F_{\rm in}$ (master-in) and average of all of out-of transit spectra is considered as $F_{\rm out}$ (master-out). 
However, this method has a problem: As mentioned before, due to orbital motion the exoplanetary RV changes and this causes the shifts of the planetary sodium features relative to the stellar spectrum, hence the signal is diluted in the process of division.
To overcome this problem, after telluric corrections and alignment of spectra, we divide each individual in-transit spectrum by the master-out spectrum. Then, to line up the exoplanetary features, we shift each residual based on the corresponding calculated radial velocity of the exoplanet in each exposure. We finally co-add all residuals \citep[similar to][]{Wyttenbach2015, Wyttenbach2017} and normalize the result with the same method explained in Section \ref{sec:data_reduction}. We then consider the normalized residual as the transmission spectrum (see Section \ref{sec:TS} and Figure \ref{fig:inout} (middle panel). By this method we correct for the RV component in the division approach as well.



It is important to note that in high-resolution transmission spectroscopy accurate alignment of the spectra is a very important step of the analysis. In the division slight misalignments can cause systematic dips or spikes in the residuals. The precision of the alignment depends on signal-to-noise and on the wavelength sampling (or the resolution) of spectra. Simulating the scatter and the wavelength sampling of our spectra on a Gaussian or Voigt distribution, 
we obtain an uncertainty of up to 0.004 \AA\, for the line position.
This amount of mis-alignment can still be a source of variations in residuals at the location of strong lines compared to the noise level at the continuum.
Dividing two sets of out-of-transit spectra by each other is a good approach to test the alignment and the robustness of the signals in transmission spectra. Hence, we also divide the average of out-of-transit spectra in exposures 1 to 4 (F$_{out1}$) before the ingress by the master-out spectra after the egress in exposures 30 to 34 (F$_{out2}$) by each other (see Section \ref{sec:TS} and Figure \ref{fig:inout} (bottom panel). These two sets of out-of-transit spectra have similar airmass, thus it can result in a second order correction of the tellurics.
In addition, in the case of this target, we can use the interstellar sodium lines as a reference for evaluation of the alignment. The interstellar lines are constant and thus in the division are expected to result in values that are uniformly scattered around unity. Thus, if the initial alignment does not satisfy this condition, an additional shift can be applied to improve the alignment.
In each single division stage in our analysis, we manually shift each in-transit spectrum with respect to the master-out, in a way to minimize the variations of the residuals at the position of interstellar lines. The amount of shifts we apply are between 0 and $\pm$ 0.004~\AA\, (with the steps of 0.0005 \AA) which is less than the precision of the initial alignment method. However, the final outcome with this method does not considerably alter the transmission spectrum and the residuals of the interstellar features do not reduce by the additional manual shifts. In Figure \ref{fig:inout} the result of a set of manual shift trial on in-transit exposures are shown in gray beneath the red profile. In contrast, in the case of division of two sets of out-transit spectra, applying an additional shift of about 0.004~\AA\, to one of the spectra, reduces the scatter of residuals. In Figure \ref{fig:inout} (bottom panel), gray profile shows the initial division and green profile shows the division after applying the manual shifts.

\begin{figure}
    \includegraphics[width = \columnwidth]{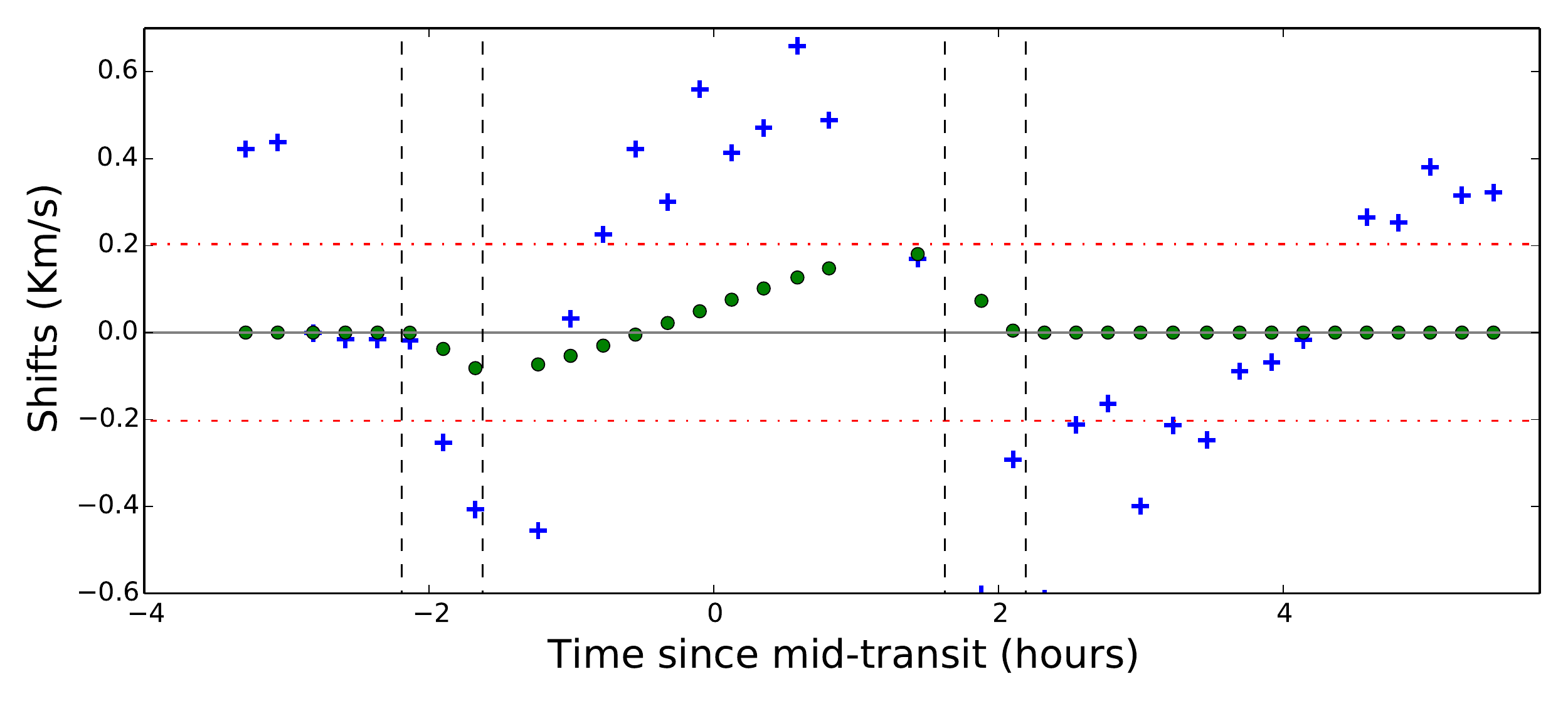}
    \caption{\label{fig:RM_oplot} RM model vs. spectral mis-alignments of sodium region. Blue crosses, show the misalignment in the sodium lines region (already shown in Figure \ref{fig:AM_shift}. Green field circles show the calculated RM effect. Red dash-dotted line shows the alignment precision of the data ($\pm$ 0.004 \AA). RM effect is below the precision of the alignment.}
\end{figure}

We also need to think of any effect that causes a difference between the alignment of interstellar and the stellar lines. The RM effect is a possible source that affects the stellar lines but do not influence the interstellar lines. Based on the literature values of the orbital parameters \citep[][]{Anderson2011}, we simulated a model of the RM effect in transit of WASP-17b and reduced it from the values of the spectral mis-alignment introduced in Section \ref{sec:data_reduction}. The RM model is shown in Figure \ref{fig:RM_oplot} on top of spectral shift values. The largest difference that the RM effect can cause is up to 0.003 \AA\, which is again below the precision of the initial alignment with our method. The removal of the RM effect from the shifts does not considerably affect the residual outcome (similar to the manual shifts in the second panel). Hence we continue the work while ignoring this effect.

\subsection{Transmission spectrum of the division approach}
\label{sec:TS}

Figure \ref{fig:inout} summarizes the results of the division approach analysis.
The master-out spectrum (or the stellar spectrum) is shown in the top panel, where the stellar and interstellar sodium lines can be clearly seen.
The second panel shows the radial velocity corrected transmission spectrum of WASP-17b. The exoplanetary sodium absorption by D1 is visible at ~5896~\AA\, and the absorption at D2 is placed at ~5890~\AA.  
 Fitting a Gaussian in each sodium feature in the transmission spectrum, yields $\overline{\sigma}_{\mathrm{Na}}$ = (0.085 $\pm$ 0.034)~\AA\, and $\overline{A}_{\mathrm{Na}}$ = (1.3 $\pm$ 0.6)\%. The error of each Gaussian fit is estimated through a MCMC procedure, where the error bar on each residual data point is equal to standard deviation of continuum region in the residuals scaled by the stellar flux.
 
 Another relatively large feature we see in the residuals (middle panel) is the scatter of data at the location of interstellar sodium lines. We note that the flux values at stellar and interstellar lines are low, thus the S/N of data at the core of these lines are lower compared to the continuum. Hence, larger scatter is inevitably present in the residuals at the location of these lines (further discussed in Section \ref{sec:discuss_interstellar}). 

 The division of two sets of out-of-transit spectra by each other are also shown as complementary and comparative information in the last panels of Figure \ref{fig:inout}. Naturally, we expect no feature at the sodium line positions in this profile. We see that the scatter of the points is relatively large at stellar and interstellar sodium line positions.
 However, the number out-of-transit exposures that build up this profile are less than the profile in the second panel by about a factor of 2 and,
 on average they are taken at larger airmass and lower S/N compared to the exposures that build up the profile in the second panel. Hence, even in the continuum the scatter is larger compared to the second panel. Since the features in these residuals at any of the sodium lines sodium lines, do not exceed 3$\sigma$ we consider the green profile  as uniform scatter around unity.
 
 Further discussion on evaluating the robustness of the exoplanetary absorption feature is in Section \ref{sec:discussion}. Our observed transmission spectrum is then compared to a planetary atmospheric model in Section \ref{sec:atm_model_division}.

\subsection{Residuals of the division}
\label{sec:discuss_interstellar}
The residuals in Figure \ref{fig:inout} (middle panel) shows relatively larger variations at the interstellar line positions. As discussed in Section \ref{sec:TS} these features are related to the lower S/N at these lines. As can be seen in the top panel of the figure, the interstellar D$_{2}$ line has the largest depth among all four sodium lines; the feature of the residual related to this line is also more pronounced. However, other reasons such as mis-alignments smaller than 0.004~\AA\, or changes of spectral line resolutions due to Earth's atmospheric effects can create such features.
Looking at the un-binned residuals of \citet[][]{Wyttenbach2017} for the case of WASP-49b, we see larger variations at the small interstellar sodium lines in their data as well. Compared to that work, interstellar lines in our case are stronger. In addition, MIKE is not as stable as HARPS and the resolution of our instrument is smaller by about a factor of two and thus the spectral alignment precision is less by about a factor of two. In any case, a complete removal of these features is not possible with this dataset and we continue the analysis assuming that features of residual spectrum at the position of the stellar sodium line are caused by the excess absorption of the exoplanetary atmosphere. 


\section{Discussion}
\label{sec:discussion}

\subsection{Excess light curve vs. division approach}

In our light curve approach we do not directly measure any light curve depth. Instead we estimate the shape of the exoplanetary sodium Gaussian profile. In the division approach we obtain the transmission spectrum in the sodium region, where the shape of exoplanetary sodium line is directly visible.

Table \ref{tbl:best-fit-values}, shows the values of widths and amplitudes of the fitted Gaussian at different passbands. The robustness of the measurements decreases with the increase of passbands, thus we consider the measurement at 0.75 \AA\, as the best measurement. In Table \ref{tbl:measure-compare} we compare the $\sigma_{Na}$ and the A$_{Na}$ of exoplanetary sodium Gaussian obtained in light curve approach, with those obtained in division approach. Within the error bars, the values are consistent with each other.

Our measured error bars on the exoplanetary feature are quite large. A comparison of the standard deviation of the data outside the transit with the expected signal at the sodium lines shows that the detection cannot reach confidence levels larger than 2$\sigma$ and this illustrates the under-estimation of the error bars in previous analysis of the same data set.

\begin{table}
\centering
\caption {sigma and amplitude of the exoplanetary Gaussian in light curve approach Vs. sigma and amplitude in division approach}
\label{tbl:measure-compare}
\begin{tabular}{ l c c}
  \hline\hline
            & $\sigma_{\mathrm{Na}}$ (\AA\,) &  A$_{\mathrm{Na}}$ \\
   \hline
    Light curve approach & 0.128 $\pm$ 0.078 & 0.017 $\pm$ 0.009\\
   \hline
    Division approach  & 0.085 $\pm$ 0.034  & 0.013 $\pm$ 0.006\\
   \hline
 \end{tabular}
 \end{table}

\begin{figure}
    \includegraphics[width = \columnwidth]{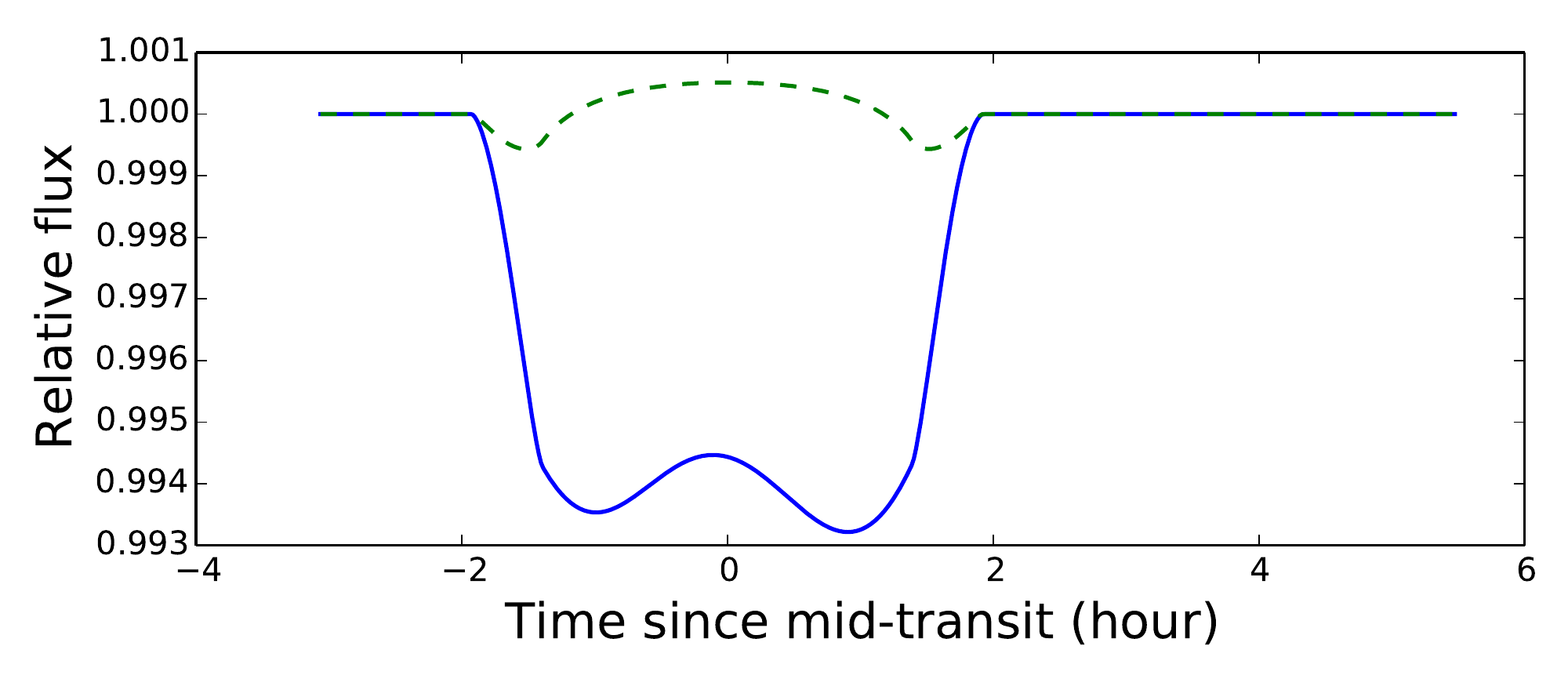}
    \caption{\label{fig:LD_vs_RV} The differential limb-darkening model component (dashed curve) vs. the radial velocity model component (solid curve) for the integration passband of 0.75~\AA.}
\end{figure}

\noindent We note that in the division approach we do not apply any differential limb-darkening correction on the spectra and their residuals directly. However, investigation of the model components in Figure \ref{fig:LD_vs_RV} indicates that the LD model is about an order of magnitude smaller than the exoplanetary absorption effects in a F-type star. Thus the strength of this effect is already within the error bars and can be ignored in the division approach.


\subsection{Comparison to previous measurements}

\subsubsection{Comparison of absorption signals}
\begin{table*}
\centering
\caption {Estimated signal in per cent, for both light curve (LC) and Division (Div.) approaches, compared to the measurments by Zhou et al. (2012), Wood et al. (2011), Sing et al. (2016) and Sedaghati et al. (2016). The last row roughly shows the expected absorption signal, considering the relation of  $f(x) = \frac{1}{x}$ for the absorption as a function of passband and using our measurement at 0.75~\AA\, (LC) as the narrowest absorption band.}
\label{tbl:signal}
\normalsize{
\begin{tabular}{ l c c c c c c}
  \hline\hline
    Passband (\AA\,) & 0.75  &  1 & 1.5 & 3 & 5 & 50\\
   \hline
    This work (LC) & 0.74 $\pm$ 0.54 & 0.55 $\pm$ 0.43 & 0.40 $\pm$ 0.29 & 0.19 $\pm$ 0.15 & & \\
   \hline
    This work (Div.) & 0.46 $\pm$ 0.29 & 0.34 $\pm$ 0.20 & 0.22 $\pm$ 0.12 & 0.12 $\pm$ 0.06 &  & \\
   \hline\hline
   Zhou et al. &    &     &   0.58 $\pm$ 0.13  &    &   &  \\
   \hline
   Wood et al. & 1.46 $\pm$ 0.017 &   & 0.55 $\pm$ 0.13 &  0.49 $\pm$ 0.09 &  & \\
   \hline
    Sing et al. &  &   &    &  & 0.33 $\pm$ 0.18 &  \\
    \hline
    Sedaghati et al. &  &   &    &  &   &  0.10 $\pm$ 0.09  \\
    \hline
    1/$x$   &  0.74 $\pm$ 0.54   & 0.55 $\pm$ 0.41 & 0.37 $\pm$ 0.026 & 0.18 $\pm$ 0.13 & 0.11 $\pm$ 0.08  & 0.01 $\pm$ 0.01\\
    \hline
 \end{tabular}
 }
 \end{table*}

In order to be able to compare our measurements with other work, we average out the exoplanetary line flux in different passbands in both approaches. The results are shown in Table \ref{tbl:signal} and the values are compared to measurements of \cite{Zhou2012} and \citet{Wood2011}, obtained from high-resolution analysis.
These measurements are consistent with each other within the error bars. The source of the difference between them could be in RV component correction and in the different approach in measurement of the signal. For example, \cite{Zhou2012} measure directly the depth of the fitted light curve while we measure the average of the flux in the fitted exoplanetary sodium line Gaussian. By looking at Figure \ref{fig:best-res} at 0.75 \AA\, and roughly measuring the depth of the modeled light curve, we achieve a similar result to \cite{Zhou2012}.

We additionally calculate the absorption signal presented in the low-resolution transmission spectrum of WASP-17b by \citet{Sing2016} as well as the upper limit on the non-detection of sodium in the transmission spectrum obtained by \citet{Sedaghati2016}. The results are also shown in Table \ref{tbl:signal}. To compute the values, we measure the difference in the $R_{\rm P}$/$R_{\rm S}$ value between the data point at the wavelength of sodium and the continuum level, and we also covert the radius ratios ($R_{\rm P}$/$R_{\rm S}$) to the flux ratios ($F_{\rm in}$/$F_{\rm out}$) to unify the units. Assuming a relation of $f(x) = \frac{1}{x}$ for the absorption as a function of passband and considering that all the exoplanetary absorption signal is coming from the line center at 0.75~\AA\, passband, we compute the values listed in the last row of the table. 
If the exoplanetary sodium line wings are broader than the high-resolution passbands, then the low-resolution bin size covers the exoplanetary feature more completely.
Comparison of our high-resolution expected values to the low-resolution measurements at 5~\AA\, and 50~\AA\, passbands/bin size, shows that the values in low-resolution observations are larger but the results are consistent with each other within the error-bars. Influence of stellar activity can be another source of variability between the measurements of different epochs.

\subsubsection{Offset between the center of the stellar and planetary sodium lines}
By looking at the sodium line center in the top and middle panel in Figure \ref{fig:inout}, we recognize an offset of $\sim$ 0.15\AA\, (7.5 kms$^{-1}$) between sodium residuals and the center of stellar sodium line. \citet{Wyttenbach2015} has seen a similar shift in the Na signal of HD~189733b.
We must emphasis that the alignment of the exoplanetary features before co-adding the residuals is highly dependent on the accuracy and precision of the calculated radial velocities of the exoplanet relative to its host star. 
In calculation of the radial velocities, changing the values of orbital parameters, i.e., semi-major axis and period, even within their error bars affects this offset. For example, considering a 2-$\sigma$ upper limit on the semi-major axis calculated by \citet{Anderson2010} or 1-$\sigma$ upper limit by \citet{Southworth2012}, the maximum offset that we measure reaches about 6 kms$^{-1}$ which corresponds to about 0.12~\AA\,.


\subsection{Atmospheric physical properties}
\label{sec:physical_parameters}

\begin{figure*}
    \centering
    \includegraphics[width =0.75\textwidth]{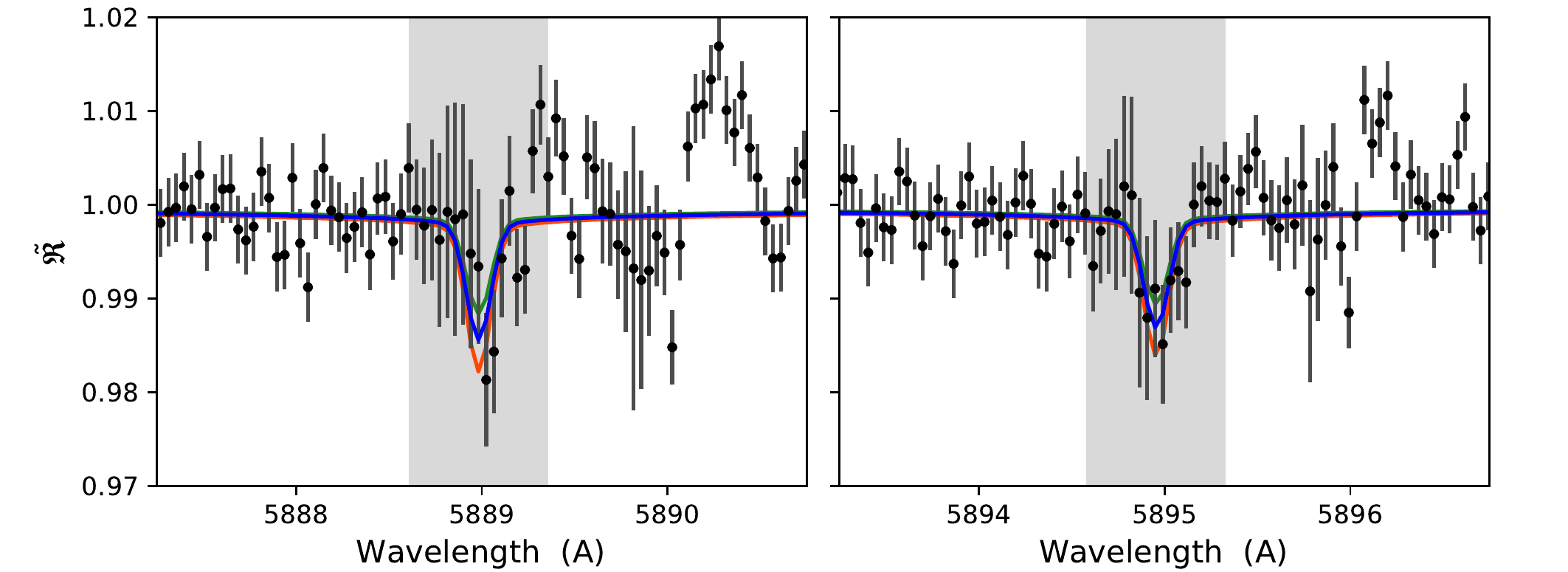}
    \caption{High-resolution spectrum (black points with $1\sigma$ error bars) and best-fitting model (solid blue line) around the sodium doublet lines for the retrieval at solar Na abundance. The shaded area denotes the wavelength region used to constrain the models. The green and red lines denote the model spectrum if $R_0$ were five percent smaller or larger, respectively.}
    \label{fig:atmospheric_model}
\end{figure*}

\begin{figure*}
    \centering
    \includegraphics[width =0.8\textwidth]{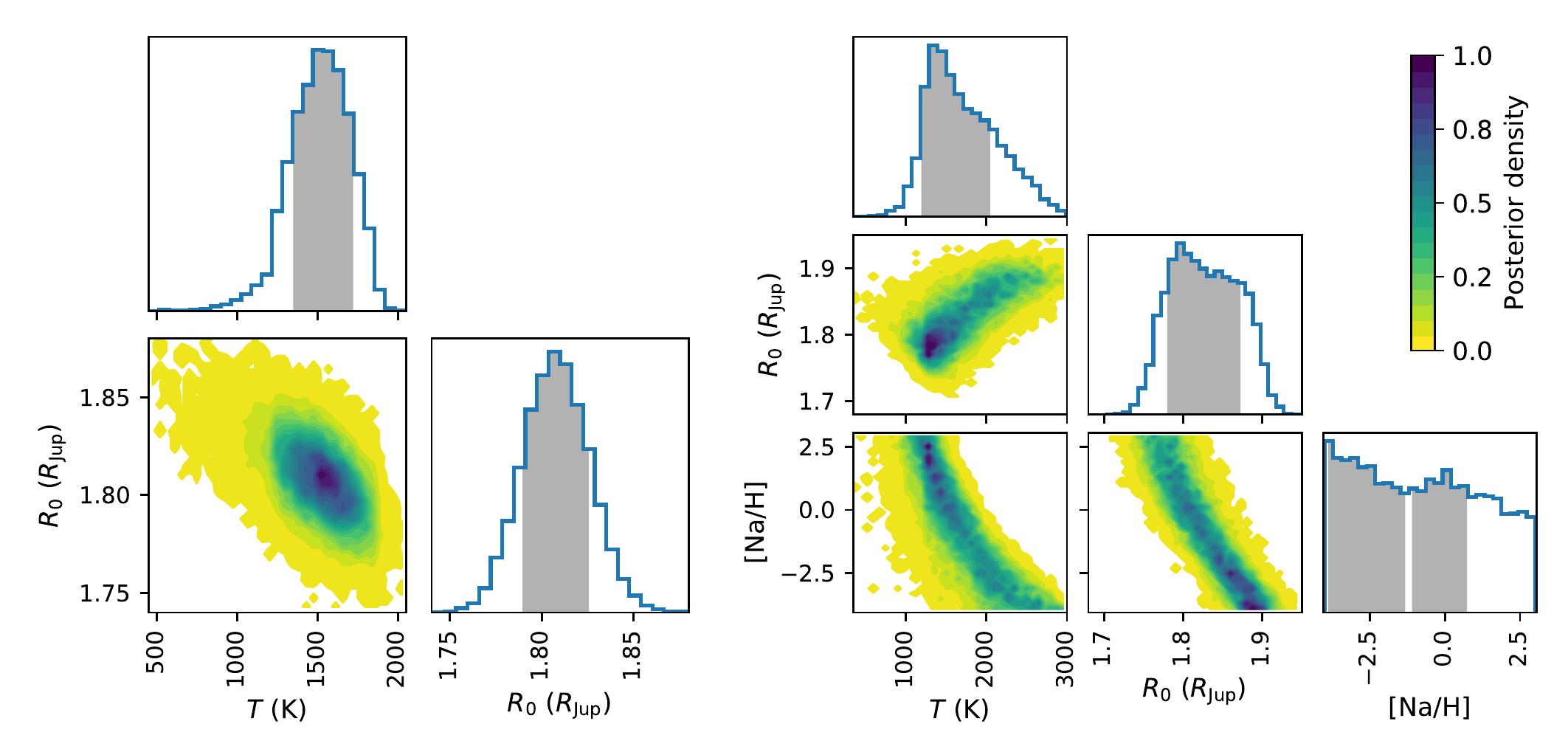}
    \caption{\label{fig:posterior} Pairwise and marginal MCMC posterior distributions for the atmospheric retrievals with fixed (left) and free (right) Na abundance.  The shaded area in the marginal posterior histograms denote the 68\% highest-probability-density region of the posteriors.} 
\end{figure*}



\subsubsection{Comparison to atmospheric models: Division approach}
\label{sec:atm_model_division}

To model the WASP-17b high-resolution transmission spectrum and obtain a physical description of the data, we use the
open-source Python Radiative Transfer in a Bayesian framework (Pyrat-Bay\footnote{\href{http://pcubillos.github.io/pyratbay}{http://pcubillos.github.io/pyratbay}\\ In addition, a Reproducible Research Compendium for the Pyrat-Bay analysis is available at:\\
\href{https://github.com/pcubillos/KhalafinejadEtal2018\_WASP17b}
     {https://github.com/pcubillos/KhalafinejadEtal2018\_WASP17b}}, Cubillos et al., in prep.), based on the Bayesian Atmospheric Radiative Transfer
package \citep{Blecic2016phdThesis, Cubillos2016phdThesis}.

Due to the wavelength normalization (Section 3.3), high-resolution
transmission spectra do not constrain the planet-to-star radius ratio
(as is the case for lower-resolution transmission spectroscopy).  For
the narrow wavelength range covered by our observations, the spectra
are dominated by the strong sodium doublet lines embedded into the
Rayleigh absorption, which sets a continuum transmission level around
the sodium lines.  Thus, this data traces the differential
transmission modulation of the sodium lines with respect to the
Rayleigh continuum.

The Pyrat-Bay model initially computes the modulation spectrum or spectrum ratio \citep{Brown2001}:
\begin{equation}
M(\lambda) = \frac{f_{\rm in}(\lambda)-f_{\rm out}(\lambda)}
                  {f_{\rm out}(\lambda)},
\label{eq:modulation}
\end{equation}
where $f_{\rm in}(\lambda)$ and $f_{\rm out}(\lambda)$ are the in- and
out-of-transit flux spectrum, respectively.  To replicate the
wavelength normalization of the high-resolution data, the code
calculates
\begin{equation}
\tilde{\mathfrak{R}} = 
   \frac{1-M(\lambda)}{1-M(\lambda_{\rm ref})} =
            \frac{f_{\rm in} (\lambda)/f_{\rm in} (\lambda_{\rm ref})}
                 {f_{\rm out}(\lambda)/f_{\rm out}(\lambda_{\rm ref})},
\label{eq:rtilde}
\end{equation}
where $\lambda_{\rm ref}$ is a wavelength far away from the sodium
lines.

To produce the transmission spectra, the Pyrat-Bay code solves the
radiative-transfer equation for an 1D atmospheric model consisting of
spherically concentric layers, in hydrostatic equilibrium.  For the
WASP-17b data, we sample the atmosphere between 100 and $10^{-15}$ bar
with 200 layers.  Our forward model incorporates opacities for H$_2$
Rayleigh scattering
\citep{LecavelierDesEtangsEtal2008aaRayleighHD189}, and the sodium
lines \citep{BurrowsEtal2000apjBDspectra}.  Other opacity sources like
collision induced absorption do not play a significant role at the
observed wavelengths as their opacities decay exponentially as one goes toward shorted wavelengths \citep[see, for example, figures in][]{AbelEtal2011jpcHydrogenCIA, AbelEtal2012jcpH2HeliumCIA}.  We also discard cloud opacities, based on
previous transmission observations of this planet
\citep{Sing2016,
  Sedaghati2016}.

For the retrieval we consider the simplified case of a solar-abundance
atmosphere with thermochemical-equilibrium compositions
\citep{BlecicEtal2016apsjTEA}.  Thus, we retrieve two atmospheric
parameters, the atmospheric temperature ($T$, as an isothermal
profile) and the radius of the planet ($R_0$) at a reference pressure
$p_0=0.1$~bar (necessary to solve the differential hydrostatic
equation).

To explore the parameter space, Pyrat-Bay uses the
differential-evolution Markov-chain Monte Carlo (MCMC) algorithm
\citep[][]{CubillosEtal2017apjRednoise}, constrained by the
high-resolution data in the 3 half-width at half maximum region around each sodium line. We
also fit the modulation (Eq. \ref{eq:modulation}) to the optical
broad-band transit depth $0.01524 \pm 0.00027$
\citep{Sedaghati2016}.

As part of our modeling approach, we solve the hydrostatic-equilibrium equation
to relate the altitude and pressure profiles of the model: $r=r(p)$.  Since this
is a first-order differential equation, we need a 'boundary condition'
of the form $R_0 = r(p_0)$ to obtain the particular solution of the
equation.  The typical procedure is to either fix $R_0$ or $p_0$, and
then find their $p_0$ or $R_0$ counterpart (respectively) that fits
the observations.  The selection of the fixed reference point is
arbitrary (we choose $p_0=0.1$~bar in this article).
The challenge is that transmission observations do not directly
constrain the pair ${R_0, p_0}$, but rather allow for degenerate
solutions to this problem \citep[see,
e.g.,][]{Griffith2014rsptaDegenerateSolutions}.  Thus, the nature of
the problem require us to include $R_0$ as a free parameter of the
atmospheric model.  We include the broad-band constraint to break down
one of the degeneracies of the atmospheric model, although other
correlations still remain.

We adopt two configurations for the retrieval.  An initial run assumes
a fixed solar Na composition of 1.7 parts per million
\citep{AsplundEtal2009araSolarComposition}, letting free the
temperature and reference radius.  A second run, lets the sodium
abundance vary as a free parameter along with the other two parameters.  All model parameters have uniform
priors, and thus, the observations are the main driver of the MCMC
posterior distribution.

Figures \ref{fig:atmospheric_model} shows our atmospheric model over the high-resolution transmission data (residuals) 
and Figure \ref{fig:posterior} shows the bestfit model and the
parameter posteriors, respectively. When adopting a fixed solar sodium abundance, we retrieve an atmospheric
temperature of $T = 1550_{-200}^{+170}$~K, which is consistent to
$1\sigma$ with the planet's equilibrium temperature ($1770 \pm 35$~K),
and a reference radius of $R_0 = 1.81 \pm 0.02$ $R_{\rm Jup}$.
As expected, the retrieval with free sodium abundances has broader posterior distributions.  The degeneracy of solutions lead to strongly correlated posteriors.  The sodium abundance is unconstrained within our chosen exploration domain of $10^{-4}$--$10^{3}$ solar abundances.  The retrieved temperature is $T = 1250_{-70}^{+800}$~K, while the reference radius ranges between 1.78--1.88 $R_{\rm Jup}$.

\subsubsection{Comparison to atmospheric models: Excess light curve approach}
\label{sec:atm_model_LC}
\begin{figure}
    \includegraphics[width =\columnwidth]{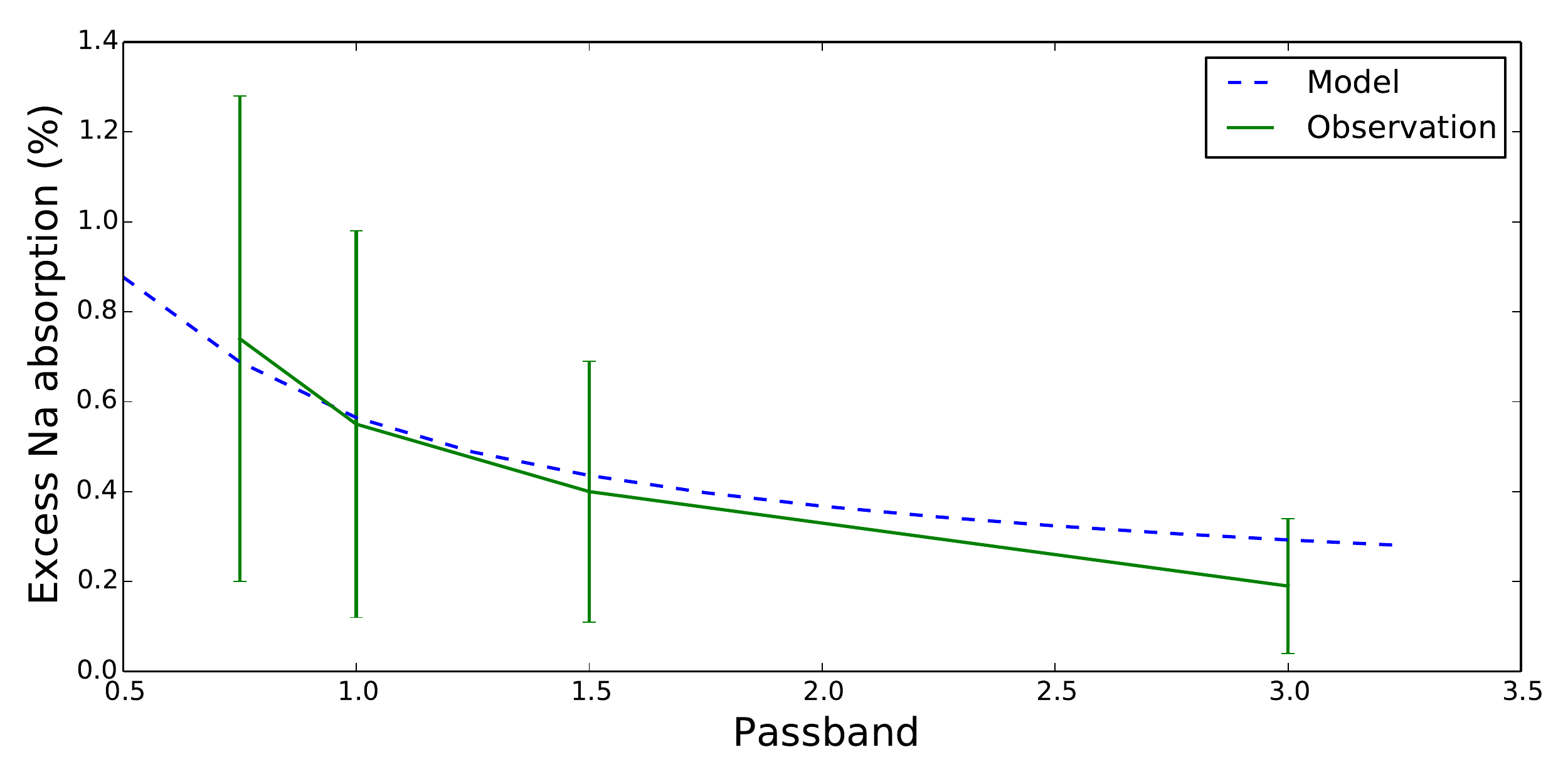}
    \caption{Atmospheric model vs. observation in the excess light curve approach. The dashed blue line is the result of averaging atmospheric model at the sodium absorption features in different passbands. The green data, attached to each other by the solid line, are the result of our ecxess light curve measurements in four different passbands.}
    \label{fig:model_obs_passband}
\end{figure}

The model that fits the residuals at the sodium lines in division approach, can be used for comparison to the excess light curve absorption signals. For this purpose after applying the instrumental broadening, we change the sampling of the models to match the sampling of the observations and then average out the model in different passbands the same way that we estimate the absorption signal in the observations. The result of this comparison is shown in Figure \ref{fig:model_obs_passband}. As the figure shows the model and the observational data are coherent.
Here we note that it is not practical to  apply an independent atmospheric model fit to this approach, since the current model is already well within the large error bars.

\section{Summary and conclusions}
\label{sec:conc}
The aim of this work is the atmospheric characterization of the inflated hot Jupiter, WASP-17b, using transmission spectroscopy method in narrow wavelength bands centered on the sodium lines.
We used 37 high-resolution spectra taken during a single transit of the target, using MIKE instrument on Magellan Telescopes.
Our analysis consists of three main sections: (1) Investigation of the stellar activity through chromospheric lines of H$_{\alpha}$ and \ion{Ca}{ii} IRT, (2) investigation of the exoplanetry sodium absorption through the excess light curves in narrow pass-bands and (3) through the division of in-transit by the out-of-transit spectra.

We detect no strong chromospheric line absorption or strong stellar variability during the observations.
In the excess light curves we detect tentative signatures of exoplanetary sodium with the Gaussian width of (0.128 $\pm$ 0.078)~\AA\, and Gaussian amplitude of (1.7 $\pm$ 0.9)\%. Through the the division approach the measured Gaussian width and amplitudes are (0.085 $\pm$ 0.034)~\AA\, and (1.3 $\pm$ 0.6)\% respectively.
Finally, in order to extract some of the physical properties we compare our results with the planetary atmospheric models. Our conclusions are listed below:

\begin{itemize}
    \item Our re-analysis of this data set suggests under-estimation of the uncertainties in the exoplanetary absorption signal measured by \citet[][]{Zhou2012}. This is originated in the way of interpreting the excess light curve.
    \item High-precision alignment (> 0.004~\AA\,) and high signal-to-noise (> 100) of the spectra are crucial in achieving a robust ground-based transmission spectrum through a single transit observation.
    \item Comparing our measurements with the Pyrat-Bay retrieval model, for WASP-17b we constrained an atmospheric temperature of $1550_{-200}^{+170}$~K, and a reference radius at 0.1 bar of $1.81 \pm 0.02$ $R_{\rm Jup}$
    \item During high-resolution transmission spectroscopy on sodium lines, simultaneous high-resolution spectral observation of the stellar activity indicators, such as \ion{Ca}{ii} H \& K, H$_{\alpha}$ and \ion{Ca}{ii} IRT, reveals possible effects of stellar flaring activity on transmission spectra. 
    \item We have developed a framework for high-resolution transmission spectroscopy in narrow passbands inside atomic lines. This framework can be applied on a higher quality data, for achieving more robust exoplanetary signals at higher confidence level. 
\end{itemize}

\begin{acknowledgements}
S. Khalafinejad would like to firstly thank M. Holman for providing the required funding for visiting the Center for Astrophysics (CfA), where we could establish this work.
In addition, we acknowledge \emph{Deut\-sche For\-schungs\-ge\-mein\-schaft (DFG),\/} in the framework of RTG 1351 and MIN faculty of Hamburg University for providing the second part of funding of this project.
C. von Essen acknowledges funding for the Stellar Astrophysics Centre, provided by The Danish National Research Foundation (Grant DNRF106).
We would also like to thank B. Fuhrmeister, S. Czesla, L. Fossati, M. Lindle, M. Guedel, J. Hoeijmakers, F. Yan and N. Espinoza for useful scientific discussions on various topics related to exoplanetray atmospheric studies.
We additionally appreciate the National Collaborative Research Infrastructure Strategy of the Australian Federal Government, who supported the access to the Magellan Telescopes.
We are also grateful to contributors to
Numpy \citep{vanderWaltEtal2011numpy},
SciPy \citep{JonesEtal2001scipy},
Matplotlib \citep{Hunter2007ieeeMatplotlib},
the Python Programming Language, and the free and open-source
community. Finally, we thank the anonymous referee for the useful comments which resulted in improvement of the contents of the paper.
\end{acknowledgements}


\bibliography{bib}

\begin{thebibliography}{70}
\expandafter\ifx\csname natexlab\endcsname\relax\def\natexlab#1{#1}\fi

\bibitem[{{Abel} {et~al.}(2011){Abel}, {Frommhold}, {Li}, \&
  {Hunt}}]{AbelEtal2011jpcHydrogenCIA}
{Abel}, M., {Frommhold}, L., {Li}, X., \& {Hunt}, K. L.~C. 2011, The Journal of
  Physical Chemistry A, 115, 6805, pMID: 21207941

\bibitem[{{Abel} {et~al.}(2012){Abel}, {Frommhold}, {Li}, \&
  {Hunt}}]{AbelEtal2012jcpH2HeliumCIA}
{Abel}, M., {Frommhold}, L., {Li}, X., \& {Hunt}, K. L.~C. 2012, \jcp, 136,
  044319

\bibitem[{{Albrecht}(2008)}]{Albrecht2008Thes}
{Albrecht}, S. 2008, PhD thesis, Leiden Observatory, Leiden University,
  P.O.~Box 9513, 2300 RA Leiden, The Netherlands

\bibitem[{{Allart} {et~al.}(2017){Allart}, {Lovis}, {Pino}, {Wyttenbach},
  {Ehrenreich}, \& {Pepe}}]{Allart2017}
{Allart}, R., {Lovis}, C., {Pino}, L., {et~al.} 2017, \aap, 606, A144

\bibitem[{{Anderson} {et~al.}(2010){Anderson}, {Hellier}, {Gillon}, {Triaud},
  {Smalley}, {Hebb}, {Collier Cameron}, {Maxted}, {Queloz}, {West}, {Bentley},
  {Enoch}, {Horne}, {Lister}, {Mayor}, {Parley}, {Pepe}, {Pollacco},
  {S{\'e}gransan}, {Udry}, \& {Wilson}}]{Anderson2010}
{Anderson}, D.~R., {Hellier}, C., {Gillon}, M., {et~al.} 2010, \apj, 709, 159

\bibitem[{{Anderson} {et~al.}(2011){Anderson}, {Smith}, {Lanotte}, {Barman},
  {Collier Cameron}, {Campo}, {Gillon}, {Harrington}, {Hellier}, {Maxted},
  {Queloz}, {Triaud}, \& {Wheatley}}]{Anderson2011}
{Anderson}, D.~R., {Smith}, A.~M.~S., {Lanotte}, A.~A., {et~al.} 2011, \mnras,
  416, 2108

\bibitem[{{Andretta} {et~al.}(2005){Andretta}, {Bus{\`a}}, {Gomez}, \&
  {Terranegra}}]{Andretta2005}
{Andretta}, V., {Bus{\`a}}, I., {Gomez}, M.~T., \& {Terranegra}, L. 2005, \aap,
  430, 669

\bibitem[{{Asplund} {et~al.}(2009){Asplund}, {Grevesse}, {Sauval}, \&
  {Scott}}]{AsplundEtal2009araSolarComposition}
{Asplund}, M., {Grevesse}, N., {Sauval}, A.~J., \& {Scott}, P. 2009, \araa, 47,
  481

\bibitem[{{Bayliss} {et~al.}(2010){Bayliss}, {Winn}, {Mardling}, \&
  {Sackett}}]{Bayliss2010}
{Bayliss}, D.~D.~R., {Winn}, J.~N., {Mardling}, R.~A., \& {Sackett}, P.~D.
  2010, \apjl, 722, L224

\bibitem[{{Bento} {et~al.}(2014){Bento}, {Wheatley}, {Copperwheat}, {Fortney},
  {Dhillon}, {Hickman}, {Littlefair}, {Marsh}, {Parsons}, \&
  {Southworth}}]{Bento2014}
{Bento}, J., {Wheatley}, P.~J., {Copperwheat}, C.~M., {et~al.} 2014, \mnras,
  437, 1511

\bibitem[{{Blecic}(2016)}]{Blecic2016phdThesis}
{Blecic}, J. 2016, ArXiv e-prints

\bibitem[{{Blecic} {et~al.}(2016){Blecic}, {Harrington}, \&
  {Bowman}}]{BlecicEtal2016apsjTEA}
{Blecic}, J., {Harrington}, J., \& {Bowman}, M.~O. 2016, \apjs, 225, 4

\bibitem[{{Bourrier} {et~al.}(2015){Bourrier}, {Lecavelier des Etangs}, \&
  {Vidal-Madjar}}]{Bourrier2015}
{Bourrier}, V., {Lecavelier des Etangs}, A., \& {Vidal-Madjar}, A. 2015, \aap,
  573, A11

\bibitem[{{Brown}(2001)}]{Brown2001}
{Brown}, T.~M. 2001, \apj, 553, 1006

\bibitem[{{Burrows} {et~al.}(2000){Burrows}, {Marley}, \&
  {Sharp}}]{BurrowsEtal2000apjBDspectra}
{Burrows}, A., {Marley}, M.~S., \& {Sharp}, C.~M. 2000, \apj, 531, 438

\bibitem[{{Bus{\`a}} {et~al.}(2007){Bus{\`a}}, {Aznar Cuadrado}, {Terranegra},
  {Andretta}, \& {Gomez}}]{Busa2007}
{Bus{\`a}}, I., {Aznar Cuadrado}, R., {Terranegra}, L., {Andretta}, V., \&
  {Gomez}, M.~T. 2007, \aap, 466, 1089

\bibitem[{{Cauley} {et~al.}(2017{\natexlab{a}}){Cauley}, {Redfield}, \&
  {Jensen}}]{Cauley2017}
{Cauley}, P.~W., {Redfield}, S., \& {Jensen}, A.~G. 2017{\natexlab{a}}, ArXiv
  e-prints

\bibitem[{{Cauley} {et~al.}(2017{\natexlab{b}}){Cauley}, {Redfield}, \&
  {Jensen}}]{Cauley2017b}
{Cauley}, P.~W., {Redfield}, S., \& {Jensen}, A.~G. 2017{\natexlab{b}}, \aj,
  153, 81

\bibitem[{{Cauley} {et~al.}(2016){Cauley}, {Redfield}, {Jensen}, \&
  {Barman}}]{Cauley2016}
{Cauley}, P.~W., {Redfield}, S., {Jensen}, A.~G., \& {Barman}, T. 2016, \aj,
  152, 20

\bibitem[{{Cessateur} {et~al.}(2010){Cessateur}, {Kretzschmar}, {Dudok de Wit},
  \& {Boumier}}]{Cessateur2010}
{Cessateur}, G., {Kretzschmar}, M., {Dudok de Wit}, T., \& {Boumier}, P. 2010,
  \solphys, 263, 153

\bibitem[{{Charbonneau} {et~al.}(2002){Charbonneau}, {Brown}, {Noyes}, \&
  {Gilliland}}]{Charbonneau2002}
{Charbonneau}, D., {Brown}, T.~M., {Noyes}, R.~W., \& {Gilliland}, R.~L. 2002,
  \apj, 568, 377

\bibitem[{{Chmielewski}(2000)}]{Chmielewski2000}
{Chmielewski}, Y. 2000, \aap, 353, 666

\bibitem[{{Cincunegui} {et~al.}(2007){Cincunegui}, {D{\'{\i}}az}, \&
  {Mauas}}]{Cincunegui2007}
{Cincunegui}, C., {D{\'{\i}}az}, R.~F., \& {Mauas}, P.~J.~D. 2007, \aap, 469,
  309

\bibitem[{{Cubillos} {et~al.}(2017){Cubillos}, {Harrington}, {Loredo}, {Lust},
  {Blecic}, \& {Stemm}}]{CubillosEtal2017apjRednoise}
{Cubillos}, P., {Harrington}, J., {Loredo}, T.~J., {et~al.} 2017, \aj, 153, 3

\bibitem[{{Cubillos}(2016)}]{Cubillos2016phdThesis}
{Cubillos}, P.~E. 2016, ArXiv e-prints

\bibitem[{{Czesla} {et~al.}(2009){Czesla}, {Huber}, {Wolter}, {Schr{\"o}ter},
  \& {Schmitt}}]{Czesla2009}
{Czesla}, S., {Huber}, K.~F., {Wolter}, U., {Schr{\"o}ter}, S., \& {Schmitt},
  J.~H.~M.~M. 2009, \aap, 505, 1277

\bibitem[{{Czesla} {et~al.}(2015){Czesla}, {Klocov{\'a}}, {Khalafinejad},
  {Wolter}, \& {Schmitt}}]{Czesla2015}
{Czesla}, S., {Klocov{\'a}}, T., {Khalafinejad}, S., {Wolter}, U., \&
  {Schmitt}, J.~H.~M.~M. 2015, \aap, 582, A51

\bibitem[{{Czesla} {et~al.}(2017){Czesla}, {Salz}, {Schneider}, {Mittag}, \&
  {Schmitt}}]{Czesla2017}
{Czesla}, S., {Salz}, M., {Schneider}, P.~C., {Mittag}, M., \& {Schmitt},
  J.~H.~M.~M. 2017, \aap, 607, A101

\bibitem[{{Foreman-Mackey} {et~al.}(2013){Foreman-Mackey}, {Conley},
  {Meierjurgen Farr}, {Hogg}, {Long}, {Marshall}, {Price-Whelan}, {Sanders}, \&
  {Zuntz}}]{Foreman2013}
{Foreman-Mackey}, D., {Conley}, A., {Meierjurgen Farr}, W., {et~al.} 2013,
  {emcee: The MCMC Hammer}, Astrophysics Source Code Library

\bibitem[{{Fortney} {et~al.}(2010){Fortney}, {Shabram}, {Showman}, {Lian},
  {Freedman}, {Marley}, \& {Lewis}}]{Fortney2010}
{Fortney}, J.~J., {Shabram}, M., {Showman}, A.~P., {et~al.} 2010, \apj, 709,
  1396

\bibitem[{{Griffith}(2014)}]{Griffith2014rsptaDegenerateSolutions}
{Griffith}, C.~A. 2014, Philosophical Transactions of the Royal Society of
  London Series A, 372, 20130086

\bibitem[{{Hauschildt} \& {Baron}(1999)}]{Hauschildt1999}
{Hauschildt}, P.~H. \& {Baron}, E. 1999, Journal of Computational and Applied
  Mathematics, 109, 41

\bibitem[{{Heng}(2016)}]{Heng2016}
{Heng}, K. 2016, \apjl, 826, L16

\bibitem[{{Hoeijmakers} {et~al.}(2015){Hoeijmakers}, {de Kok}, {Snellen},
  {Brogi}, {Birkby}, \& {Schwarz}}]{Hoeijmakers2015}
{Hoeijmakers}, H.~J., {de Kok}, R.~J., {Snellen}, I.~A.~G., {et~al.} 2015,
  \aap, 575, A20

\bibitem[{{H{\o}g} {et~al.}(2000){H{\o}g}, {Fabricius}, {Makarov}, {Urban},
  {Corbin}, {Wycoff}, {Bastian}, {Schwekendiek}, \& {Wicenec}}]{Hog2000}
{H{\o}g}, E., {Fabricius}, C., {Makarov}, V.~V., {et~al.} 2000, \aap, 355, L27

\bibitem[{{Huitson} {et~al.}(2012){Huitson}, {Sing}, {Vidal-Madjar},
  {Ballester}, {Lecavelier des Etangs}, {D{\'e}sert}, \& {Pont}}]{Huitson2012}
{Huitson}, C.~M., {Sing}, D.~K., {Vidal-Madjar}, A., {et~al.} 2012, \mnras,
  422, 2477

\bibitem[{Hunter(2007)}]{Hunter2007ieeeMatplotlib}
Hunter, J.~D. 2007, Computing In Science \& Engineering, 9, 90

\bibitem[{{Husser} {et~al.}(2013){Husser}, {Wende-von Berg}, {Dreizler},
  {Homeier}, {Reiners}, {Barman}, \& {Hauschildt}}]{Husser2013}
{Husser}, T.-O., {Wende-von Berg}, S., {Dreizler}, S., {et~al.} 2013, \aap,
  553, A6

\bibitem[{{Jensen} {et~al.}(2012){Jensen}, {Redfield}, {Endl}, {Cochran},
  {Koesterke}, \& {Barman}}]{Jensen2012}
{Jensen}, A.~G., {Redfield}, S., {Endl}, M., {et~al.} 2012, \apj, 751, 86

\bibitem[{Jones {et~al.}(2001)Jones, Oliphant, Peterson,
  {et~al.}}]{JonesEtal2001scipy}
Jones, E., Oliphant, T., Peterson, P., {et~al.} 2001, {SciPy}: Open source
  scientific tools for {Python}, [Online; accessed 2017-02-12]

\bibitem[{{Kempton} {et~al.}(2014){Kempton}, {Perna}, \& {Heng}}]{Kempton2014}
{Kempton}, E.~M.-R., {Perna}, R., \& {Heng}, K. 2014, \apj, 795, 24

\bibitem[{{Khalafinejad} {et~al.}(2017){Khalafinejad}, {von Essen},
  {Hoeijmakers}, {Zhou}, {Klocov{\'a}}, {Schmitt}, {Dreizler}, {Lopez-Morales},
  {Husser}, {Schmidt}, \& {Collet}}]{Khalafinejad2017}
{Khalafinejad}, S., {von Essen}, C., {Hoeijmakers}, H.~J., {et~al.} 2017, \aap,
  598, A131

\bibitem[{{Klocova} {et~al.}(2017){Klocova}, {Czesla}, {Khalafinejad},
  {Wolter}, \& {Schmitt}}]{Klocova2017}
{Klocova}, T., {Czesla}, S., {Khalafinejad}, S., {Wolter}, U., \& {Schmitt},
  J.~H.~M.~M. 2017, ArXiv e-prints

\bibitem[{{Kopal}(1950)}]{Kopal1950}
{Kopal}, Z. 1950, Harvard College Observatory Circular, 454, 1

\bibitem[{{Lecavelier Des Etangs} {et~al.}(2008){Lecavelier Des Etangs},
  {Pont}, {Vidal-Madjar}, \&
  {Sing}}]{LecavelierDesEtangsEtal2008aaRayleighHD189}
{Lecavelier Des Etangs}, A., {Pont}, F., {Vidal-Madjar}, A., \& {Sing}, D.
  2008, \aap, 481, L83

\bibitem[{{Mandell} {et~al.}(2013){Mandell}, {Haynes}, {Sinukoff},
  {Madhusudhan}, {Burrows}, \& {Deming}}]{Mandell2013}
{Mandell}, A.~M., {Haynes}, K., {Sinukoff}, E., {et~al.} 2013, \apj, 779, 128

\bibitem[{{Mart{\'{\i}}nez-Arn{\'a}iz}
  {et~al.}(2010){Mart{\'{\i}}nez-Arn{\'a}iz}, {Maldonado}, {Montes}, {Eiroa},
  \& {Montesinos}}]{Martinez2010}
{Mart{\'{\i}}nez-Arn{\'a}iz}, R., {Maldonado}, J., {Montes}, D., {Eiroa}, C.,
  \& {Montesinos}, B. 2010, \aap, 520, A79

\bibitem[{{Moehler} {et~al.}(2014){Moehler}, {Modigliani}, {Freudling},
  {Giammichele}, {Gianninas}, {Gonneau}, {Kausch}, {Lan{\c c}on}, {Noll},
  {Rauch}, \& {Vinther}}]{Moehler2014}
{Moehler}, S., {Modigliani}, A., {Freudling}, W., {et~al.} 2014, \aap, 568, A9

\bibitem[{{Morley} {et~al.}(2015){Morley}, {Fortney}, {Marley}, {Zahnle},
  {Line}, {Kempton}, {Lewis}, \& {Cahoy}}]{Morley2015}
{Morley}, C.~V., {Fortney}, J.~J., {Marley}, M.~S., {et~al.} 2015, \apj, 815,
  110

\bibitem[{{Nortmann}(2015)}]{NortmannThesis}
{Nortmann}, L. 2015, PhD thesis, der Georg-August-Universit\"at G\"ottingen

\bibitem[{{Oshagh} {et~al.}(2013){Oshagh}, {Santos}, {Boisse}, {Bou{\'e}},
  {Montalto}, {Dumusque}, \& {Haghighipour}}]{Oshagh2013}
{Oshagh}, M., {Santos}, N.~C., {Boisse}, I., {et~al.} 2013, \aap, 556, A19

\bibitem[{{Oshagh} {et~al.}(2014){Oshagh}, {Santos}, {Ehrenreich},
  {Haghighipour}, {Figueira}, {Santerne}, \& {Montalto}}]{Oshagh2014}
{Oshagh}, M., {Santos}, N.~C., {Ehrenreich}, D., {et~al.} 2014, \aap, 568, A99

\bibitem[{{Pont} {et~al.}(2008){Pont}, {Knutson}, {Gilliland}, {Moutou}, \&
  {Charbonneau}}]{Pont2008}
{Pont}, F., {Knutson}, H., {Gilliland}, R.~L., {Moutou}, C., \& {Charbonneau},
  D. 2008, \mnras, 385, 109

\bibitem[{{Reiners}(2012)}]{Reiners2012}
{Reiners}, A. 2012, Living Reviews in Solar Physics, 9, 1

\bibitem[{{Salz} {et~al.}(2016){Salz}, {Schneider}, {Czesla}, \&
  {Schmitt}}]{Salz2016}
{Salz}, M., {Schneider}, P.~C., {Czesla}, S., \& {Schmitt}, J.~H.~M.~M. 2016,
  \aap, 585, L2

\bibitem[{{Seager} \& {Sasselov}(2000)}]{Seager2000}
{Seager}, S. \& {Sasselov}, D.~D. 2000, \apj, 537, 916

\bibitem[{{Sedaghati} {et~al.}(2016){Sedaghati}, {Boffin}, {Je{\v
  r}abkov{\'a}}, {Garc{\'{\i}}a Mu{\~n}oz}, {Grenfell}, {Smette}, {Ivanov},
  {Csizmadia}, {Cabrera}, {Kabath}, {Rocchetto}, \& {Rauer}}]{Sedaghati2016}
{Sedaghati}, E., {Boffin}, H.~M.~J., {Je{\v r}abkov{\'a}}, T., {et~al.} 2016,
  \aap, 596, A47

\bibitem[{{Sedaghati} {et~al.}(2017){Sedaghati}, {Boffin}, {MacDonald},
  {Gandhi}, {Madhusudhan}, {Gibson}, {Oshagh}, {Claret}, \&
  {Rauer}}]{Sedaghati2017}
{Sedaghati}, E., {Boffin}, H.~M.~J., {MacDonald}, R.~J., {et~al.} 2017, \nat,
  549, 238

\bibitem[{{Sing} {et~al.}(2016){Sing}, {Fortney}, {Nikolov}, {Wakeford},
  {Kataria}, {Evans}, {Aigrain}, {Ballester}, {Burrows}, {Deming},
  {D{\'e}sert}, {Gibson}, {Henry}, {Huitson}, {Knutson}, {Etangs}, {Pont},
  {Showman}, {Vidal-Madjar}, {Williamson}, \& {Wilson}}]{Sing2016}
{Sing}, D.~K., {Fortney}, J.~J., {Nikolov}, N., {et~al.} 2016, \nat, 529, 59

\bibitem[{{Sing} {et~al.}(2012){Sing}, {Huitson}, {Lopez-Morales}, {Pont},
  {D{\'e}sert}, {Ehrenreich}, {Wilson}, {Ballester}, {Fortney}, {Lecavelier des
  Etangs}, \& {Vidal-Madjar}}]{Sing2012}
{Sing}, D.~K., {Huitson}, C.~M., {Lopez-Morales}, M., {et~al.} 2012, \mnras,
  426, 1663

\bibitem[{{Southworth} {et~al.}(2012){Southworth}, {Hinse}, {Dominik}, {Fang},
  {Harps{\o}e}, {J{\o}rgensen}, {Kerins}, {Liebig}, {Mancini}, {Skottfelt},
  {Anderson}, {Smalley}, {Tregloan-Reed}, {Wertz}, {Alsubai}, {Bozza}, {Calchi
  Novati}, {Dreizler}, {Gu}, {Hundertmark}, {Jessen-Hansen}, {Kains},
  {Kjeldsen}, {Lund}, {Lundkvist}, {Mathiasen}, {Penny}, {Rahvar}, {Ricci},
  {Scarpetta}, {Snodgrass}, \& {Surdej}}]{Southworth2012}
{Southworth}, J., {Hinse}, T.~C., {Dominik}, M., {et~al.} 2012, \mnras, 426,
  1338

\bibitem[{{Triaud} {et~al.}(2010){Triaud}, {Collier Cameron}, {Queloz},
  {Anderson}, {Gillon}, {Hebb}, {Hellier}, {Loeillet}, {Maxted}, {Mayor},
  {Pepe}, {Pollacco}, {S{\'e}gransan}, {Smalley}, {Udry}, {West}, \&
  {Wheatley}}]{Triaud2010}
{Triaud}, A.~H.~M.~J., {Collier Cameron}, A., {Queloz}, D., {et~al.} 2010,
  \aap, 524, A25

\bibitem[{van~der Walt {et~al.}(2011)van~der Walt, Colbert, \&
  Varoquaux}]{vanderWaltEtal2011numpy}
van~der Walt, S., Colbert, S.~C., \& Varoquaux, G. 2011, Computing in Science
  \& Engineering, 13, 22

\bibitem[{{Vidal-Madjar} {et~al.}(2011){Vidal-Madjar}, {Sing}, {Lecavelier Des
  Etangs}, {Ferlet}, {D{\'e}sert}, {H{\'e}brard}, {Boisse}, {Ehrenreich}, \&
  {Moutou}}]{Vidal-Madjar2011}
{Vidal-Madjar}, A., {Sing}, D.~K., {Lecavelier Des Etangs}, A., {et~al.} 2011,
  \aap, 527, A110

\bibitem[{{Wolter} {et~al.}(2005){Wolter}, {Schmitt}, \& {van
  Wyk}}]{Wolter2005}
{Wolter}, U., {Schmitt}, J.~H.~M.~M., \& {van Wyk}, F. 2005, \aap, 435, 261

\bibitem[{{Wood} {et~al.}(2011){Wood}, {Maxted}, {Smalley}, \&
  {Iro}}]{Wood2011}
{Wood}, P.~L., {Maxted}, P.~F.~L., {Smalley}, B., \& {Iro}, N. 2011, \mnras,
  412, 2376

\bibitem[{{Wyttenbach} {et~al.}(2015){Wyttenbach}, {Ehrenreich}, {Lovis},
  {Udry}, \& {Pepe}}]{Wyttenbach2015}
{Wyttenbach}, A., {Ehrenreich}, D., {Lovis}, C., {Udry}, S., \& {Pepe}, F.
  2015, \aap, 577, A62

\bibitem[{{Wyttenbach} {et~al.}(2017){Wyttenbach}, {Lovis}, {Ehrenreich},
  {Bourrier}, {Pino}, {Allart}, {Astudillo-Defru}, {Cegla}, {Heng}, {Lavie},
  {Melo}, {Murgas}, {Santerne}, {S{\'e}gransan}, {Udry}, \&
  {Pepe}}]{Wyttenbach2017}
{Wyttenbach}, A., {Lovis}, C., {Ehrenreich}, D., {et~al.} 2017, ArXiv e-prints

\bibitem[{{Yan} \& {Henning}(2018)}]{Yan2018}
{Yan}, F. \& {Henning}, T. 2018, Nature Astronomy

\bibitem[{{Zhou} \& {Bayliss}(2012)}]{Zhou2012}
{Zhou}, G. \& {Bayliss}, D.~D.~R. 2012, \mnras, 426, 2483

\end{thebibliography}

\appendix

\section{Transmission spectra around H$_{\alpha}$ and Ca II IRT }

\begin{figure*}
    \centering
    \includegraphics[width=0.45\textwidth]{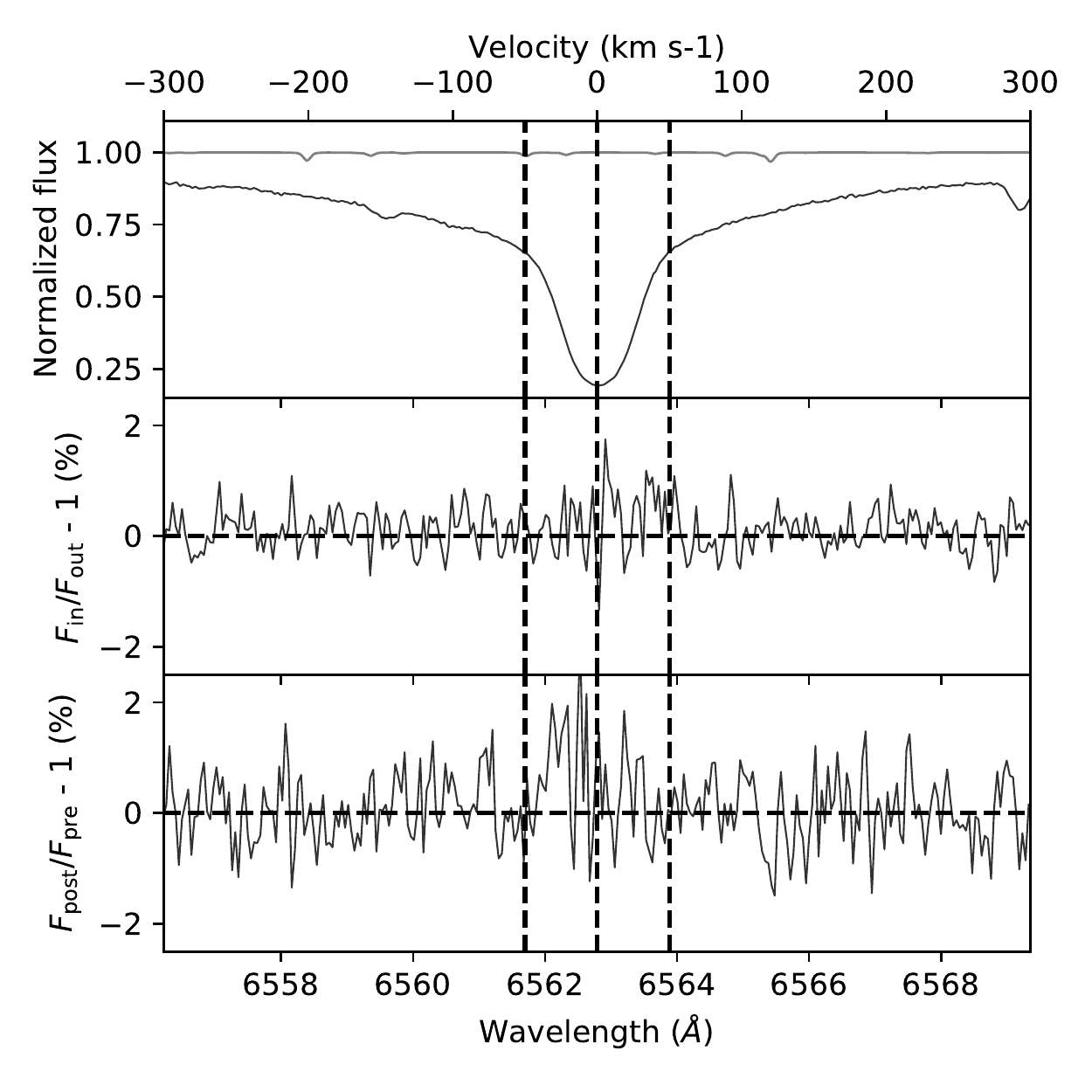}
    \includegraphics[width=0.45\textwidth]{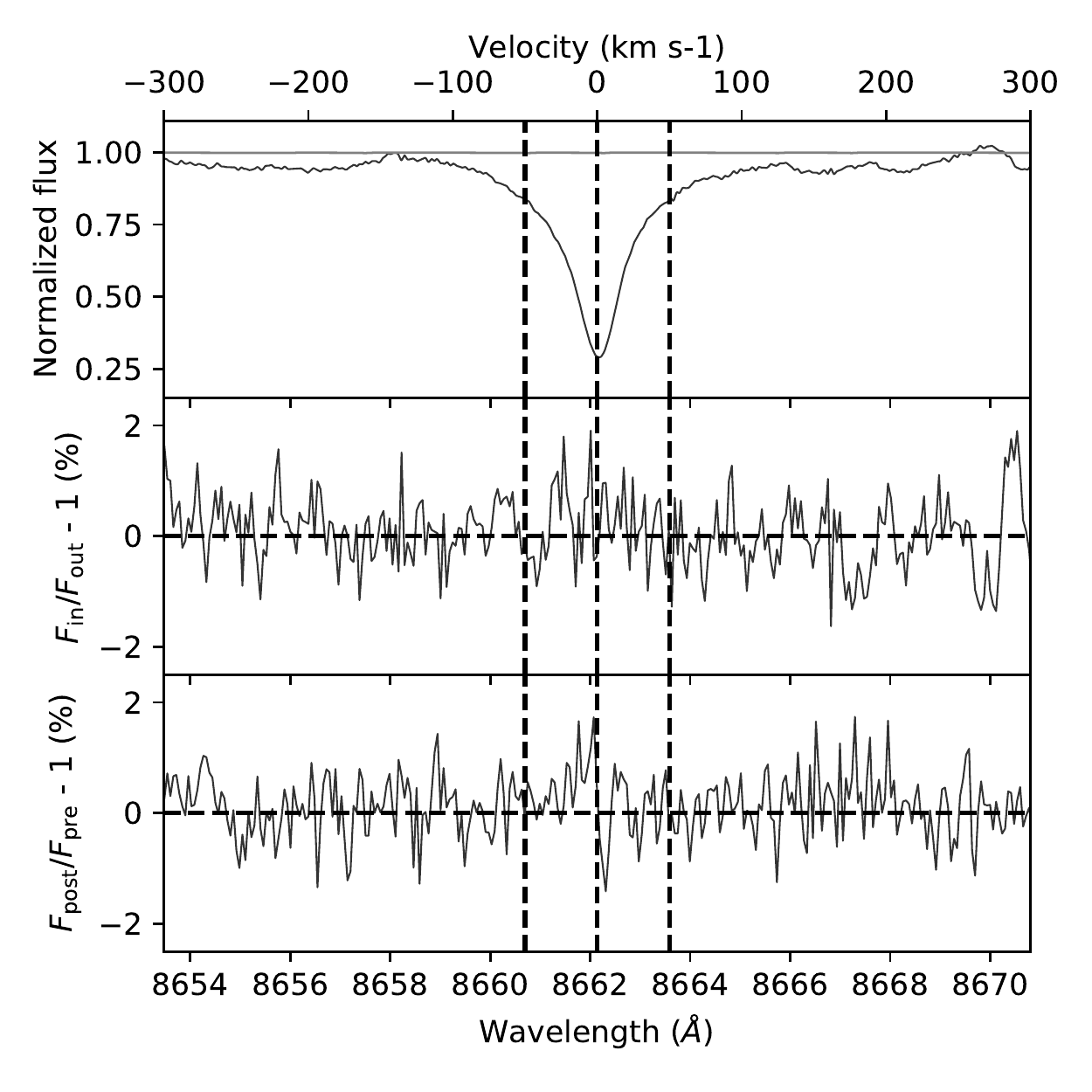}
    \caption{\textbf{left:} Transit spectrum of H$_{\alpha}$. We show in the mean stellar spectrum, the removed telluric absorption spectrum, the in- divided by out-of-transit spectrum and the post- divided by the pre-transit spectrum. There is an excess in the line core during the transit phase, most likely related to stellar activity. \textbf{Right:} The same for one of the Ca II lines.}
    \label{fig:TS_Halpha_CaIIIRT}
\end{figure*}

We confirm that the source of the variations in H$_{\alpha}$ and Ca II IRT lines is stellar activity and not the exoplanetary extended atmosphere. In Fig. \ref{fig:TS_Halpha_CaIIIRT}, we show the transmission spectra in the stellar rest frame related to the H$_{\alpha}$ and one of the Ca II IRT lines. The other two Ca II IRT lines behave similarly.

In the H$_{\alpha}$ region (Fig. \ref{fig:TS_Halpha_CaIIIRT}-left), the residuals are clean cosmic rays and telluric contamination are already removed. The strongest variation clearly occurs in the line center, and in the stellar rest frame as opposed to the planetary rest frame. We attribute this evolution to stellar activity. The IRT lines show a similar behaviour (see e.g., Fig. \ref{fig:TS_Halpha_CaIIIRT}-right). 


\section{MCMC complementary figures}

\begin{figure*}
    \centering
    \includegraphics[width = 0.7\textwidth]{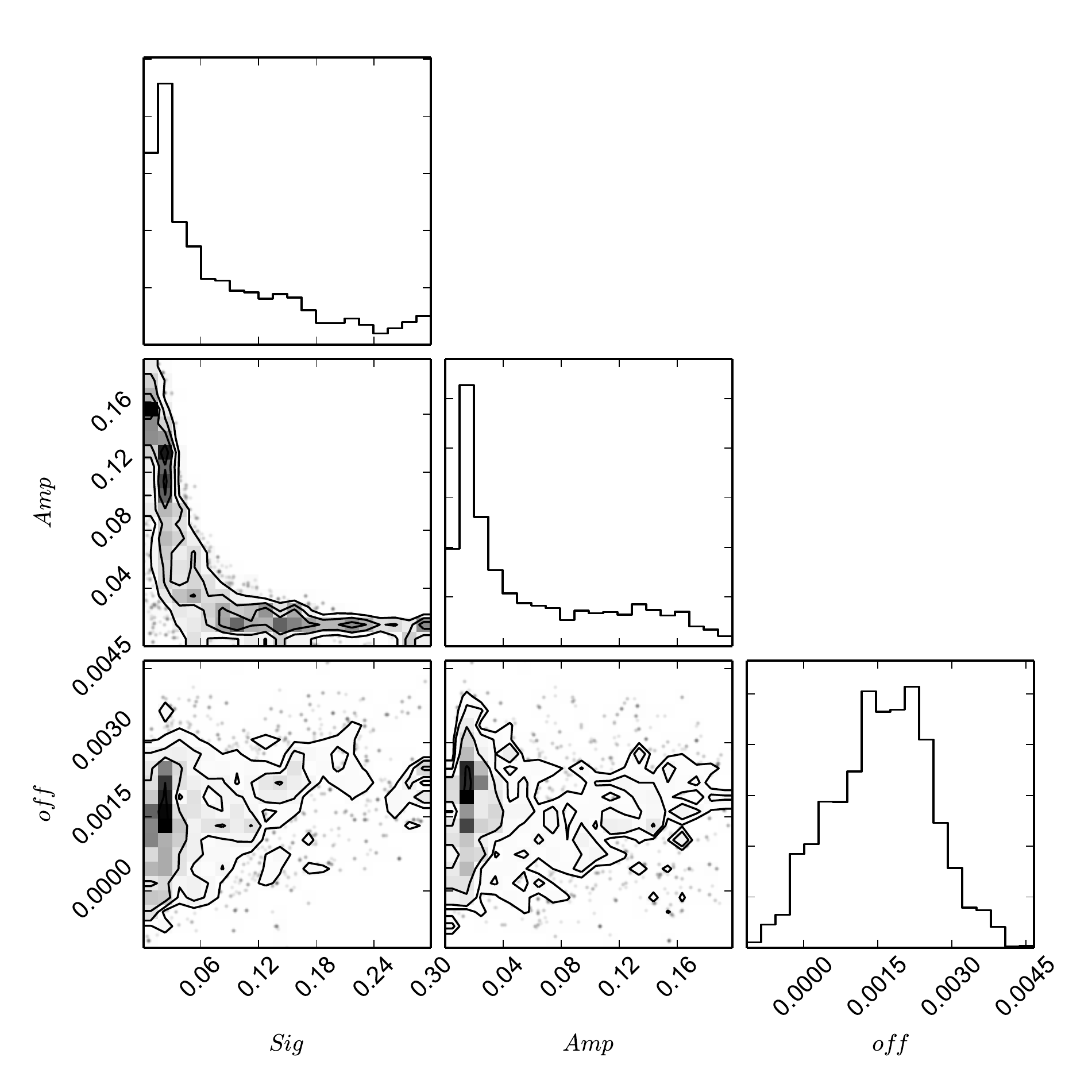}
    \caption{Posterior distributions of the model parameters fitted in this work in the shape of histograms, along with their correlation plots, for the passband of 1.5 \AA.}
    \label{fig:corner}
\end{figure*}

A sample of the MCMC posterior distribution in the excess light curve modeling is shown in Fig. \ref{fig:corner}.

\end{document}